\documentclass[bibyear]{aa}  

\usepackage{graphicx}
\usepackage{hyperref}
\hypersetup{hidelinks}
\usepackage[dvipsnames]{xcolor}
\usepackage{marvosym}
\usepackage{natbib}
\bibpunct{(}{)}{;}{a}{}{,} 

\usepackage{subcaption,booktabs}
\usepackage{comment}
\usepackage[thinc]{esdiff}
\newcommand{\msun}{{\rm M}_\odot}

\newcommand{\sevn}{\textsc{sevn}\,}
\usepackage{soul}

\defcitealias{sana2012}{S12}
\defcitealias{sb2013}{SB13}
\usepackage{txfonts}
\usepackage{pifont}
\makeatletter
\renewcommand*\aa@pageof{, page \thepage{} of \pageref*{LastPage}}
\makeatother

\newcommand{\orcidicon}[1]{\href{https://orcid.org/#1}{\includegraphics[width=11pt]{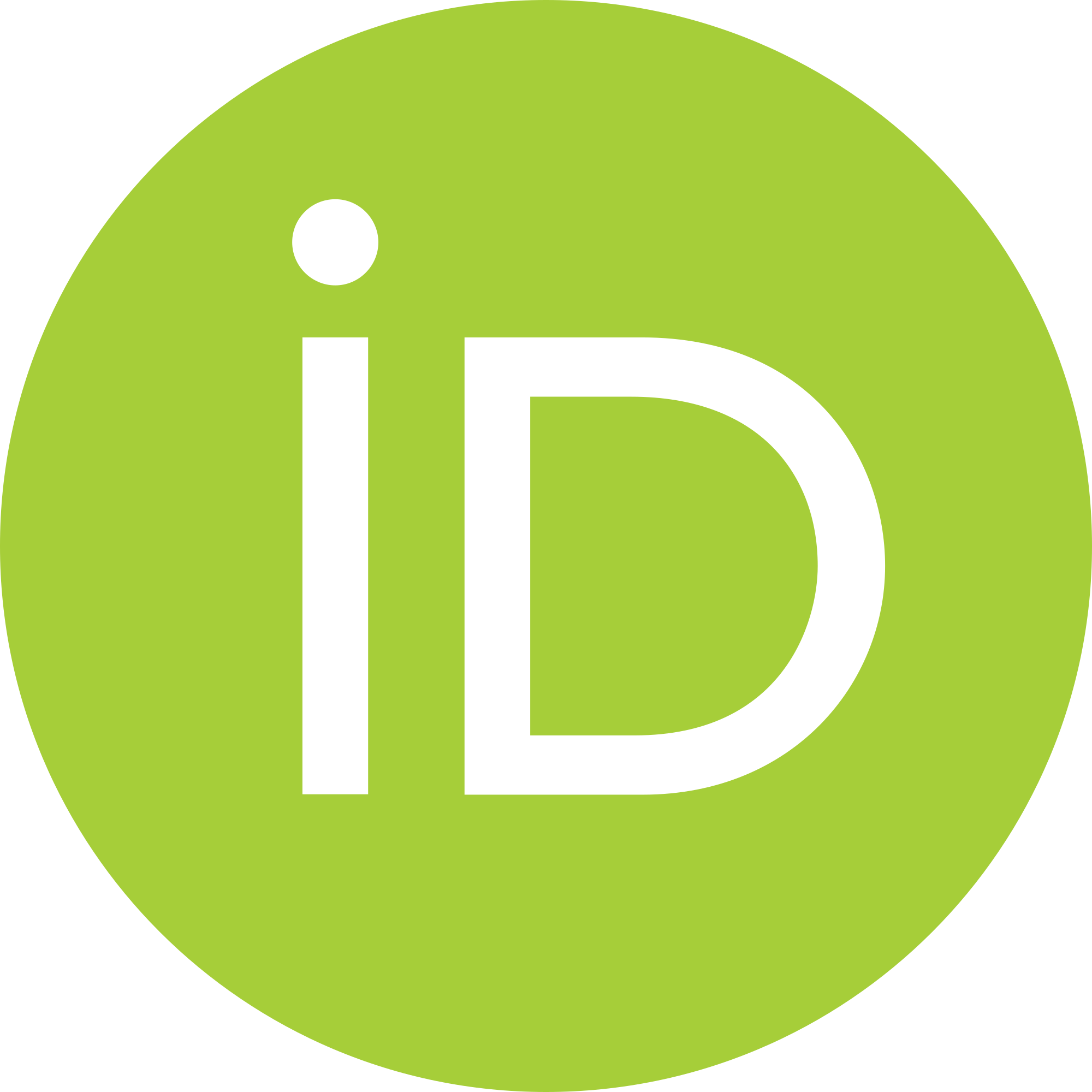}}}
\newcommand{\orcid}[1]{\href{https://orcid.org/#1}{\protect\orcidicon{#1}}}

\definecolor{seagreen}{rgb}{0.190, 0.525, 0.361}

\begin{document} 

\title{Black hole--neutron star and binary neutron star mergers from Population~III and II stars}
\authorrunning{B. Mestichelli et al.}
   \author{Benedetta Mestichelli
          \inst{1,2,3}
    \orcid{0009-0002-1705-4729} \thanks{\href{mailto:benedetta.mestichelli@gssi.it}{benedetta.mestichelli@gssi.it}}
          \and 
          Michela Mapelli\inst{3,4,5,6,7} \orcid{0000-0001-8799-2548}\thanks{\href{mailto:mapelli@uni-heidelberg.de}{mapelli@uni-heidelberg.de}}
          \and
          Filippo Santoliquido\inst{1,2}\orcid{0000-0003-3752-1400}
          \and
          Manuel Arca Sedda\inst{1,2,8}\orcid{0000-0002-3987-0519}
          \and \\
          Marica Branchesi\inst{1,8}
          \and 
          Lavinia Paiella\inst{1, 8}
          \and
          Guglielmo Costa\inst{5,7}
          \and 
          Giuliano Iorio\inst{9}
          \and Matthew Mould\inst{10,11} \orcid{0000-0001-5460-2910}
          \and Veronika Lipatova \inst{3} \orcid{0000-0002-6111-2570}
          \and Boyuan Liu \inst{3}\orcid{0000-0002-4966-7450}
          \and Ralf S.\ Klessen \inst{3,4,12,13}\orcid{0000-0002-0560-3172}}
          
          \authorrunning{B. Mestichelli et al.}
          \institute{
          $^1$Gran Sasso Science Institute (GSSI), Viale Francesco Crispi 7, 67100, L’Aquila, Italy\\
           $^2$INFN, Laboratori Nazionali del Gran Sasso, I-67100 Assergi, Italy\\
          $^3$Universit\"at Heidelberg, Zentrum f\"ur Astronomie (ZAH), Institut f\"ur Theoretische Astrophysik, Albert Ueberle Str. 2, 69120, Heidelberg, Germany\\
          $^4$Universit\"at Heidelberg, Interdisziplin\"ares Zentrum f\"ur Wissenschaftliches Rechnen, Heidelberg, Germany\\
          $^5$Physics and Astronomy Department Galileo Galilei, University of Padova, Vicolo dell'Osservatorio 3, I--35122, Padova, Italy\\
          $^6$INFN - Padova, Via Marzolo 8, I--35131 Padova, Italy\\
          $^7$INAF - Osservatorio Astronomico di Padova, Vicolo dell’Osservatorio 5, I-35122 Padova, Italy\\
          $^8$INAF Osservatorio Astronomico d'Abruzzo, Via Maggini, 64100 Teramo, Italy\\
          $^{9}$ Departament de Física Quàntica i Astrofísica, Institut de Ciències del Cosmos, Universitat de Barcelona, Martí i Franquès 1, E-08028 Barcelona, Spain\\
          $^{10}$LIGO Laboratory, Massachusetts Institute of Technology, Cambridge, MA 02139, USA\\
          $^{11}$Kavli Institute for Astrophysics and Space Research, Massachusetts Institute of Technology, Cambridge, MA 02139, USA \\
          $^{12}$Harvard-Smithsonian Center for Astrophysics, 60 Garden Street, Cambridge, MA 02138, USA \\
          $^{13}$Elizabeth S. and Richard M. Cashin Fellow at the Radcliffe Institute for Advanced Studies at Harvard University, 10 Garden Street, Cambridge, MA 02138, USA
          }
    
   \date{}

  \abstract{
  
  Population III (Pop.~III) stars are expected to be massive and to undergo minimal mass loss due to their lack of metals, making them ideal progenitors of black holes and neutron stars. Here, we investigate the formation and properties of binary neutron star (BNS) and black hole--neutron star (BHNS) mergers originating from Pop.~III stars, and compare them to their metal-enriched Population II (Pop.~II) counterparts, focusing on their merger rate densities (MRDs), primary masses, and delay times. We find that, despite the high merger efficiency of Pop.~III BNSs and BHNSs, their low star formation rate results in a MRD at least one order of magnitude lower than that of Pop.~II stars. The MRD of Pop.~III BNSs peaks at redshift $z\sim15$, attaining a value $\mathcal{R}_{\rm BNS}(z\sim15) \sim 15\,\rm Gpc^{-3}\,yr^{-1}$, while the MRD of Pop.~III BHNSs is maximum at $z\sim13$, reaching a value of $\mathcal{R}_{\rm BHNS}(z\sim13) \sim 2\,\rm Gpc^{-3}\,yr^{-1}$. Finally, we observe that the black hole masses of Pop.~III BHNS mergers have a nearly flat distribution, with a peak at $\sim20\,\msun$ and extending up to $\sim50\,\msun$. Black holes in Pop.~II BHNS mergers instead show a peak at $\lesssim15\,\msun$. We consider these predictions in light of recent gravitational-wave observations in the local Universe, finding that a Pop.~III origin is preferred relative to Pop.~II for some events.} 


   \maketitle
%

\section{Introduction}
Population III (Pop.~III) stars formed from metal-free gas at high redshift \citep{haiman1996, tegmark1997, yoshida2003}. No conclusive evidence of Pop.~III stars exists so far \citep{rydberg2013, schauer2022, larkin2023, meena2023, trussler2023}. Current models suggest that Pop.~III stars have a top-heavier initial mass function (IMF) compared to stellar populations in the local Universe \citep{sb2013,susa2014,hirano2015,jaacks2019,sharda2020,liu2020,liu2020b,wollenberg2020,chon2021,tanikawa2021,jaura2022,prole2022,klessen2023,liu2024b}. Additionally, because of their lack of metals, we expect these stars to lose a small amount of mass through stellar winds over their lifetimes, when they are not fast rotators \citep{heger2002,kinugawa2014,kinugawa2016,hartwig2016,belczynski2017,liu2021b,tanikawa2021b,tanikawa2022,tanikawa2023,costa2023,santoliquido2023,tanikawa2024}. In contrast, early Population II (Pop.~II) stars are enriched by the yields of Pop.~III stars, but still sufficiently metal-poor (with metallicities, $Z$, of up to a few $\times{}10^{-4}$, \citealt{smc2018, frebel2019}) that mass loss via line-driven stellar winds is quenched \citep{kudritzki2000, vink2001, sabhahit2023}.

Compact binary mergers from Pop.~III star progenitors could be detected by current gravitational-wave (GW) interferometers, if their delay times are sufficiently long. As a matter of fact, one of the possible scenarios for the massive binary black hole (BH) merger GW190521, forming a remnant BH of $\sim150\,\msun$, is the collapse of a Pop.~III binary. Indeed, the negligible mass loss and top-heavy IMF of Pop.~III stars might lead to the formation of a population of massive BHs and binary BHs \citep{gw190521,abbott2020a,abbott2020b,liu2020b,kinugawa2021,tanikawa2021b}. The properties of these massive compact objects have been extensively studied through a variety of numerical simulations, both in isolation \citep{tanikawa2021,tanikawa2021b,tanikawa2022,tanikawa2023, costa2023,santoliquido2023,tanikawa2024} and in star clusters \citep{sakurai2017, wang2022, mestichelli2024,liu2024,shuai2024,reinoso2025,wu2025}. Furthermore, binary BH mergers from Pop.~III stars are expected to be significant sources for third-generation interferometers, such as Cosmic Explorer \citep{ce_2019,evans2021,evans2023} and the Einstein Telescope \citep{kalogera2021,branchesi2023,santoliquido2023,santoliquido2024,etblue2025}, as these instruments will be capable of probing the high-redshift Universe, where Pop.~III star formation is expected to be prevalent.

Besides binary BHs, Pop.~III stars might also be the progenitors of binary neutron stars (BNSs) and BH--neutron star binaries (BHNSs) detectable with current and next-generation interferometers. 
By the end of the third observation run, the LIGO--Virgo--KAGRA (LVK) collaboration had observed several BHNS and BNS merger candidates \citep{gwtc1,gw190814,gw200105_gw200115,gwtc3_2023,gwtc2.1}. Furthermore, the merger GW230529, detected at the beginning of the fourth observing run, has a primary component mass of $\sim{3-6}$ M$_\odot$ \citep{gw230529}, inside the previously claimed mass gap between the maximum neutron star and minimum BH mass \citep{orosz2003,ozel2010,farr2011}. While current GW detectors only probe BNS and BHNS mergers in the nearby Universe (up to $z\sim0.2$, \citealt{gwtc3_2023}), next-generation interferometers will be sensitive to neutron star mergers in a much larger portion of the Universe, increasing the likelihood of observing BNS and BHNS mergers from Pop.~II and III star progenitors \citep{branchesi2023, gupta2024, etblue2025}.

Here, we present models of BNSs and BHNSs formed from the evolution of Pop.~III and Pop.~II binary stars. We study the distribution of their masses and delay times, and their merger rate densities. To encompass the uncertainties on Pop.~III stars, we probe different combinations of IMFs and orbital parameters, following the same approach as our previous works on binary BHs \citep{costa2023,santoliquido2023,mestichelli2024}. 
The paper is structured as follows. Section~\ref{methods} presents the initial conditions and the numerical codes used. Section~\ref{results} shows our main results. Section~\ref{discussion} provides a more detailed analysis of the properties of BHNS mergers and compares formation scenarios involving Pop.~III and Pop.~II progenitors for currently observed LVK events using Bayes factors. Finally, Section~\ref{summary} summarizes our conclusions. 

\section{Methods} \label{methods}
\subsection{
Population synthesis with \sevn}

We modeled the formation of BNSs and BHNSs with the binary population synthesis code \sevn \citep{spera2019,mapelli2020,sevn2023}, which calculates single stellar evolution by interpolating a set of precomputed single stellar tracks and describes the main binary interaction processes (mass transfer via stellar winds, Roche lobe overflow, common envelope, tides, and GW decay) by means of semi-analytic prescriptions \citep{hurley2002, sevn2023}. For the common-envelope phase, we set the efficiency parameter, $\alpha_{\rm CE}=1$, and model the envelope binding energy, $\lambda_{\rm CE}$, as was done by \cite{claeys2014}. We refer to \cite{costa2023}, \cite{santoliquido2023}, and \cite{mestichelli2024} for further details about our models.

\subsection{Compact object formation} \label{sec:sn_models}

According to the prescriptions in \sevn, depending on the final carbon-oxygen (CO) core mass ($M_{\rm CO,f}$), a star can form a white dwarf ($M_{\rm CO,f}<1.38\,\rm M_{\odot}$), undergo electron-capture supernova (SN), and form a neutron star ($1.38\le M_{\rm CO,f}<1.44\,\rm M_{\odot}$; \citealp{giacobbo2019}), or undergo core-collapse SN and form either a neutron star or a BH (CCSN; $M_{\rm CO,f}\ge 1.44\,\rm M_{\odot}$). For CCSNe, we adopted the rapid formalism by \cite{fryer2012}, according to which if $6\le M_{\rm CO,f}<7\,\rm M_{\odot}$ or $M_{\rm CO,f}\ge 11\,\rm M_{\odot}$, the star will collapse directly into a BH. As a consequence of the adopted CCSN model, neutron star masses fall between $1.19$ and $2\,\msun$, and a lower mass gap between $2$ and $5\,\msun$ naturally emerges. Given the lack of a predictive model for the mass function of neutron stars able to match the masses of Galactic neutron stars \citep{vigna_gomez2018, sgalletta2023}, we do not include neutron star masses in our analysis, for both Pop.~III and Pop.~II progenitors.

For (pulsational) pair-instability SNe, we used the fitting formulas reported by \cite{mapelli2020} and based on the simulations by \cite{woosley2017}. If the pre-SN He-core mass of a star, $M_{\rm He,f}$, is between 32 and 64 $\msun$, it undergoes pulsational pair-instability. A pair-instability SN is triggered for $64\le M_{\rm He,f} \le 135\,\msun$. Above $135\,\msun$, the star collapses directly to a BH as a consequence of photodisintegration \citep{woosley2017,spera2017,renzo2020}.

\subsection{Supernova kicks}

After a SN, compact objects receive a natal kick, which we modeled following \cite{giacobbo2020}:
\begin{equation}
    v_{\rm kick} = f_{\rm H05}\,\frac{\langle M_{\rm NS} \rangle}{M_{\rm rem}}\,\frac{M_{\rm ej}}{\langle M_{\rm ej} \rangle},
\end{equation}
with $\langle M_{\rm NS} \rangle$ and $\langle M_{\rm ej} \rangle$ computed from a population of isolated neutron stars with metallicities representative of the Milky Way, $M_{\rm rem}$ the mass of the compact object, and $M_{\rm ej}$ the ejected mass. $f_{\rm H05}$ is a number randomly drawn from a Maxwellian distribution with the one-dimensional root mean square $\sigma_{\rm kick} = 265\,\mathrm{km\,s^{-1}}$, derived from the proper motions of young Galactic pulsars \citep{hobbs2005}. This empirical formalism, while approximated, allows us to match the observations of young Galactic pulsars \citep{hobbs2005,verbunt2017}, and results in lower kicks for BHs, in accordance with observations \citep{atri2019}. In particular, it implies that BHs forming through direct collapse receive a null natal kick.

In a binary system, natal kicks affect the orbital properties, relative orbital velocity, and center of mass of the binary \citep{hurley2002}. After the kick, the orbital parameters are updated, taking into account the new relative orbital velocity and total mass of the binary. For further details, we refer to \cite{sevn2023}.

\subsection{Stellar tracks} \label{sec:star_tracks}
Our Pop.~III star tracks do not include rotation and were calculated with the stellar-evolution code \textsc{parsec} \citep{bressan2012,chen2015,costa2019,costa2025}. At the beginning of their main sequence, Pop.~III stars cannot ignite the carbon-nitrogen-oxygen (CNO) tri-cycle because of the initial lack of these elements. In order to contrast the gravitational collapse with the energy provided by the proton-proton (pp) chain, Pop.~III stars need to reach very high central temperatures. The temperature can become so high that some carbon is synthesized via a triple-$\alpha$ reaction (He burning), even during the main sequence. As a consequence, the CNO tri-cycle ignites and replaces the pp chain as the main source of energy of the star \citep{marigo2001, murphy2021}. At the end of the main sequence, Pop.~III stars have high enough central temperatures to transition smoothly to the core-He burning phase. Usually, these stellar evolution features appear at $Z<10^{-10}$ \citep{cassisi1993}. As a consequence, for Pop.~III stars we assumed an initial hydrogen abundance of $X=0.751$, He abundance of $Y=0.2485$ \citep{komatsu2011}, and metallicity of $Z=10^{-11}$ \citep{tanikawa2021,costa2023}. For Pop.~II stars we adopted a metallicity $Z=10^{-4}$.

We considered stars with a zero-age main sequence (ZAMS) mass, $m_{\rm ZAMS}$, of between 2 and 600~$\msun$. 
Stars with $2\le m_{\rm{ZAMS}}< 10\,\rm M_{\odot}$ reach the early-asymptotic giant branch phase, while stars with $m_{\rm{ZAMS}}\ge 10\,\rm M_{\odot}$ evolve until late phases of core-oxygen burning, or the beginning of the pair-instability regime. All tracks were computed with the same setup as in \cite{costa2021}, including stellar winds, a nuclear reaction network, opacity, and the equation of state. Above the convective core, we assumed a penetrative overshooting with a characteristic parameter of $\lambda_{\rm ov}=0.5$ in units of pressure scale height. For further information on the evolutionary tracks, we refer to \cite{costa2023}.

\begin{figure*}[ht]
    \centering
    \includegraphics[width=0.9\textwidth]{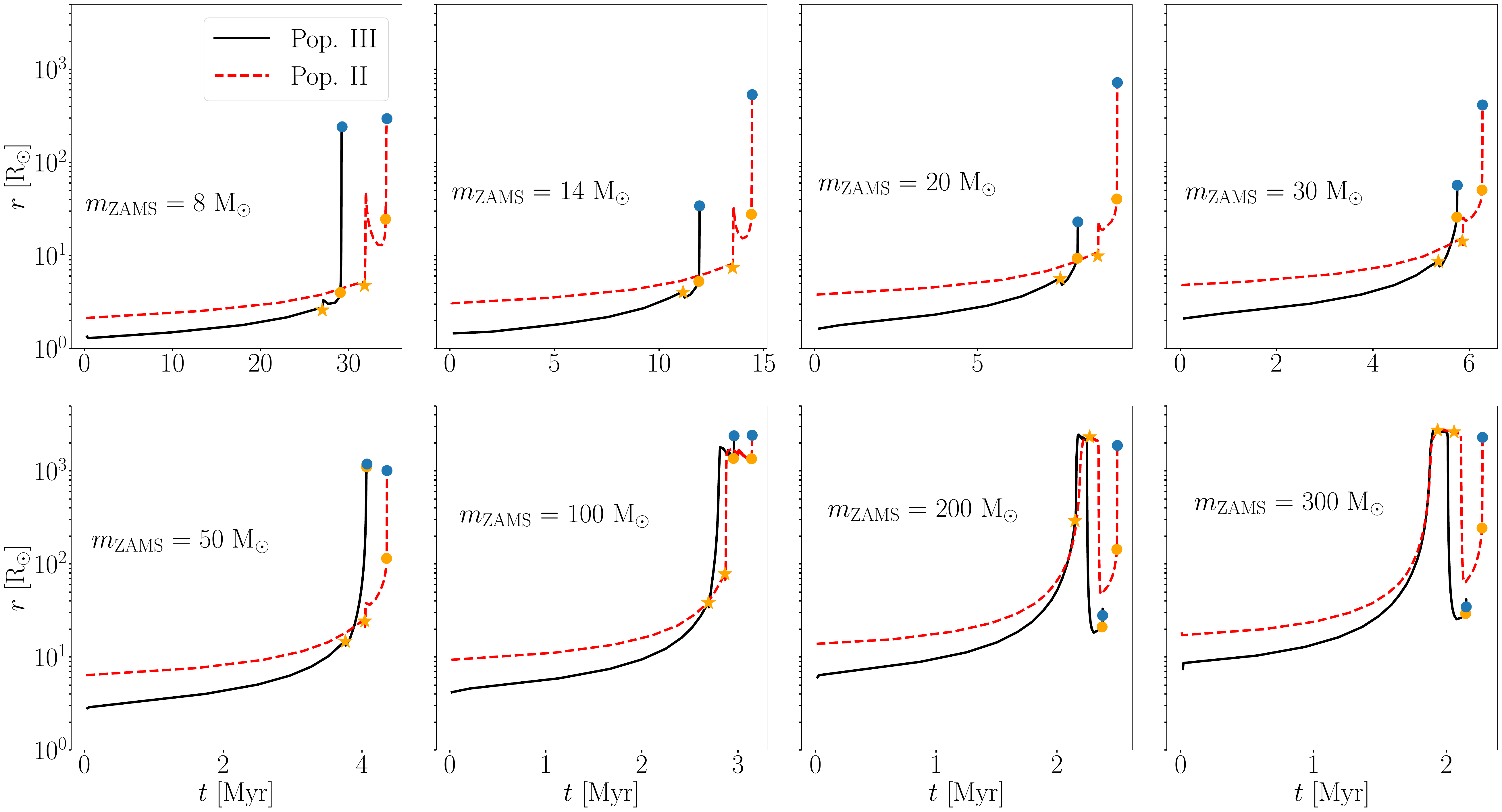}
    \caption{Evolution of stellar radii for $8 \le m_{\rm ZAMS}\leq300\,\msun$. We represent Pop.~III stars with a solid black line and Pop.~II stars with a dashed red line. The orange stars (circles) represent the beginning (end) of core-He burning. The blue circles indicate instead the pre-SN phase.}
     \label{fig:r_evol}
\end{figure*}

In Figure~\ref{fig:r_evol} we show the evolution of the radii of Pop.~III and Pop.~II stars. This figure highlights the compactness of metal-free stars compared to Pop.~II stars, especially in the mass range $8\le m_{\rm ZAMS} \le 30\,\msun$. As has already been pointed out by \cite{costa2023}, Pop.~III stars with $m_{\rm ZAMS}<100\,\msun$ begin burning He in their cores before their Pop.~II counterparts. This difference becomes less evident at higher masses. We also notice that Pop.~III stars with $m_{\rm ZAMS}>100\,\msun$ reach the pre-SN stage as blue supergiant stars, while Pop.~II stars explode as very large red supergiant stars. 

\subsection{Binary initial conditions}
The simulation sets we analyze here are the same as those presented by \cite{costa2023}. In this work, we analyze BNSs and BHNSs for the first time, whereas \cite{costa2023} focus on binary BHs. 

We assumed the following fiducial models for the IMFs and initial orbital properties of the simulated binary stars. For both Pop.~II and Pop.~III binary stars, we randomly sampled the initial mass ratios, orbital periods, and eccentricities from the distributions reported by \cite{sana2012} and obtained from observations of Galactic O-type binary stars. For Pop.~III binary stars, we randomly sampled the ZAMS masses of Pop.~III primaries from a log-flat IMF, which we were motivated to do by previous studies \citep{sb2013, susa2014, hirano2015, wollenberg2020, chon2021, tanikawa2021b, jaura2022, prole2022}. In contrast, we randomly sampled the ZAMS masses of Pop.~II primaries from a Kroupa IMF \citep{kroupa2001}. Hereafter, we refer to our fiducial model for Pop.~III and Pop.~II binary stars as log1 and kro1, respectively, for consistency with previous works \citep{costa2023,santoliquido2023,mestichelli2024}. 

In Sections~\ref{sec:mrd_bhns}, \ref{sec:mrd_bns}, and \ref{sec: bhns_primary_models}, we compare our fiducial models with additional simulations that assume a variety of initial conditions, encompassing the uncertainties on both IMF and orbital properties of Pop.~III binary stars. We describe the additional models in Appendix~\ref{ap:isolated} and summarize them in Table~\ref{table:ic_costa}. Here, we do not consider stars with $m_{\rm ZAMS}<2\,\msun$.

\begin{table}[ht!]
\centering
\caption{Initial conditions for the stellar models}
\resizebox{\columnwidth}{!}{%
\begin{tabular}{cccccc}
 \hline
 Model & $m_{\mathrm{ZAMS,1}}$ & $m_{\mathrm{ZAMS}}$ & $q$ & $P$ & $e$\\ [0.5ex] 
 \hline\hline
 kro1 & \cite{kroupa2001} & Flat in log & \citetalias{sana2012} & \citetalias{sana2012} & \citetalias{sana2012}\\ 
 kro5 & \cite{kroupa2001} & Flat in log & \citetalias{sb2013} & \citetalias{sb2013} & Thermal\\
 lar1 & \cite{larson1998} & - & \citetalias{sana2012} & \citetalias{sana2012} & \citetalias{sana2012} \\ 
 lar5 & \cite{larson1998} & - & \citetalias{sb2013} & \citetalias{sb2013} & Thermal\\
 log1 & Flat in log & - & \citetalias{sana2012} & \citetalias{sana2012} & \citetalias{sana2012}\\ 
 log2 & Flat in log & - & \citetalias{sana2012} & \citetalias{sb2013} & Thermal\\
 log3 & - & - & Sorted & \citetalias{sana2012} & \citetalias{sana2012} \\
 log4 & - & - & \citetalias{sb2013} & \citetalias{sana2012} & Thermal \\
 log5 & Flat in log & - & \citetalias{sb2013} & \citetalias{sb2013} & Thermal\\
 top1 & Top heavy & - & \citetalias{sana2012} & \citetalias{sana2012} & \citetalias{sana2012} \\ 
 top5 & Top heavy & Flat in log & \citetalias{sb2013} & \citetalias{sb2013} & Thermal \\
 \hline
\end{tabular}%
}
\tablefoot{Column 1: Name of the model. Column 2: IMF of the primary star. Column 3: ZAMS mass of the overall stellar population. Columns 4, 5, and 6: initial distributions for the mass ratio, $q$, period, $P$, and eccentricity, $e$. The acronyms \citetalias{sana2012} and \citetalias{sb2013} refer to \cite{sana2012} and \cite{sb2013}, respectively.}
\label{table:ic_costa}
\end{table}

\subsection{Formation channels} 
\label{sec:sevn_form_chan}
To interpret the properties of the simulated BNS and BHNS mergers, we refer to the six formation channels introduced by \cite{sevn2023} and defined as follows. Channel 1 involves stable mass transfer before the formation of the first compact object, followed by at least one common-envelope phase. In channel 2, the binary evolves only through stable mass transfer episodes. In both channels 3 and 4, the binary undergoes at least one common-envelope phase before the formation of the first compact object. In particular, in channel 3, the companion star retains an H-rich envelope at the time of formation of the first compact object. In channel 4, instead, the companion has been stripped of its envelope when the first compact remnant forms. 
The two remaining channels refer to systems that do not interact at all (channel 0), or interact only after the formation of the first compact object (channel 5). A summary of these formation channels can be found in Table~\ref{table:descr_form_chan}.

\begin{table}[ht!]
\renewcommand{\arraystretch}{1.5} 
\centering
\caption{Formation channels of merging compact binaries}
\begin{tabular}{c p{0.8\columnwidth}} 
 \hline
 Channel & Description \\ [0.5ex] 
 \hline\hline
 0 & No interaction \\ 
 1 & SMT before first remnant + CE after first remnant \\ 
 2 & SMT only \\ 
 3 & CE before first remnant + secondary 
 with H-rich envelope at formation of first remnant \\ 
 4 & CE before first remnant + secondary 
 stripped of envelope at formation of first remnant \\ 
 5 & Interaction only after formation of first remnant \\ 
 \hline
\end{tabular}
\tablefoot{Column 1: Number of the channel. Column 2: Brief description of the formation channel. Here SMT stands for stable mass transfer, and CE stands for common envelope.}
\label{table:descr_form_chan}
\end{table}

\subsection{Merger rate density} \label{sec:mrd_methods}

We estimated the merger rate density (MRD) of compact binaries using the semi-analytic code \textsc{cosmo}$\mathcal{R}$\textsc{ate} \citep{santoliquido2020,santoliquido2021,santoliquido2023}, which combines catalogs of simulated compact binary mergers with a model of the metal-dependent cosmic star formation rate density. Specifically, the MRD in the comoving frame was computed as
\begin{equation}
\label{eq:mrd}
    \mathcal{R}(z) = \int_{z_{{\rm{max}}}}^{z}\left[\int_{Z_{{\rm{min}}}}^{Z_{{\rm{max}}}} \,{}\psi{(z',Z)}\,{} 
    \mathcal{F}(z',z,Z) \,{}{\rm{d}}Z\right]\,{} \frac{{{\rm d}t(z')}}{{\rm{d}}z'}\,{}{\rm{d}}z',
\end{equation}
where $\psi(z', Z)$ is the metallicity-dependent cosmic star formation rate density at redshift $z'$ and metallicity $Z$, and ${{\rm{d}}t(z')}/{{\rm{d}}z'} = H_{0}^{-1}\,{}(1+z')^{-1}\,{}[(1+z')^3\Omega_{M}+ \Omega_\Lambda{}]^{-1/2}$, where $H_0$ is the Hubble parameter and $\Omega_M$ and $\Omega_\Lambda$ are the matter and energy density, respectively. We adopted here the values of the cosmological parameters from \cite{aghanim2020}. 

The term $\mathcal{F}(z',z, Z)$ in Equation~\ref{eq:mrd} is given by
\begin{equation}
\label{eq:f_mrd}
\mathcal{F}(z',z, Z) = \frac{1}{M_{\ast{}}(z',Z)}\,{}\frac{{\rm{d}}\mathcal{N}(z',z, Z)}{{\rm{d}}t(z)},
\end{equation}
where ${{\rm{d}}\mathcal{N}(z',z, Z)/{\rm{d}}}t(z)$ is the rate of BNS or BHNS mergers from stars with metallicity $Z$ that form at redshift $z'$ and merge at redshift $z$, and $M_\ast(z',Z)$ is the initial total stellar mass that forms at redshift $z'$ with metallicity $Z$. We obtained ${{\rm{d}}\mathcal{N}(z',z, Z)/{\rm{d}}}t(z)$ directly from our simulations.

Owing to the lack of direct observations of Pop.~III stars, numerous models have been proposed for their $\psi(z',Z)$ (e.g., \citealp{jaacks2019,skinner2020,liu2020}). We adopted the star formation history derived from the semi-analytic code \textsc{a-sloth}, developed by \cite{hartwig2022} and \cite{magg2022}, which tracks individual Pop.~III and Pop.~II stars and is calibrated against a range of observables from both the local and high-redshift Universe. For comparison, we also estimated the MRD of BNSs and BHNSs formed from early Pop.~II stars. In this case, we adopted the star formation rate density of Pop.~II stars ($Z\sim10^{-5}-10^{-2}\,\rm Z_{\odot}$) by \cite{liu2025}, obtained adopting a revised version of the semi-analytic code \textsc{a-sloth} \citep{hartwig2024}. This version is calibrated to reproduce the metallicity--stellar mass--star formation rate relation of high-redshift galaxies in recent observations of the James Webb Space Telescope \citep{sarkar2025}. 

\begin{figure}[ht]
    \centering
    \includegraphics[width=0.9\columnwidth]{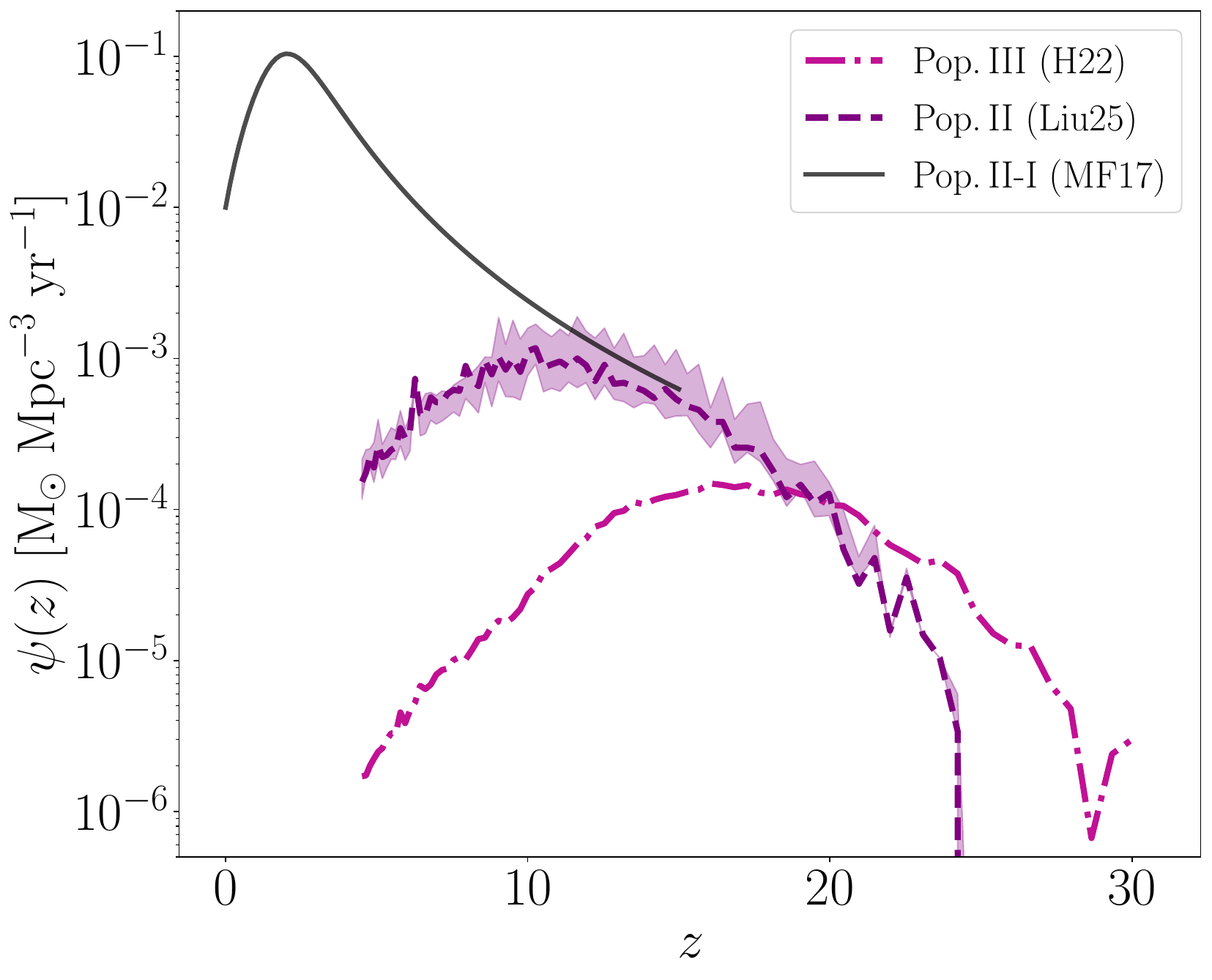}
    \caption{Star formation rate density of Pop.~III (dash-dotted magenta line, \citealp{hartwig2022}) and Pop.~II (dashed purple line, \citealp{liu2025}) stars. The shaded purple region denotes the uncertainty range associated with Pop.~II star formation history. As a comparison, the solid black line shows the star formation rate density of Pop.~II-I stars by \cite{madau2017}.}
    \label{fig:sfrd}
\end{figure}

The described star formation rate densities, $\psi(z',Z)$, are reported in Fig.~\ref{fig:sfrd}. The adopted models for $\psi(z',Z)$ predicted by \textsc{a-sloth} are valid for $z>4.5$. Therefore, it is assumed that Pop.~III and early Pop.~II star formation at later epochs is negligible. This is supported by the finding that their star formation rate density peaks at $z \sim 10 - 15$, and has dropped by a factor of 5 - 50 by $z\sim4.5$, because of metal enrichment and reionization. We also show the uncertainty range associated with Pop.~II star formation, reflecting variations in SN-driven outflows and assumptions about the IMF \citep{liu2025}.

\section{Results} \label{results}

\subsection{Black hole---neutron star binaries (BHNSs)} 
\label{sec:bhns_mergers}

In this Section, we describe the properties of BHNS mergers from Pop.~III and Pop.~II binary stars. We first discuss the impact of the simulated Pop.~III models on the MRD of BHNSs and make a comparison with the MRD of Pop.~II BHNSs. We then explore the distributions of delay times, primary masses, and mass ratios of BHNSs for the fiducial models of Pop.~III and Pop.~II star progenitors.

\subsubsection{Merger rate density} 
\label{sec:mrd_bhns}

\begin{figure}[ht]
    \centering
    \includegraphics[width=0.9\columnwidth]{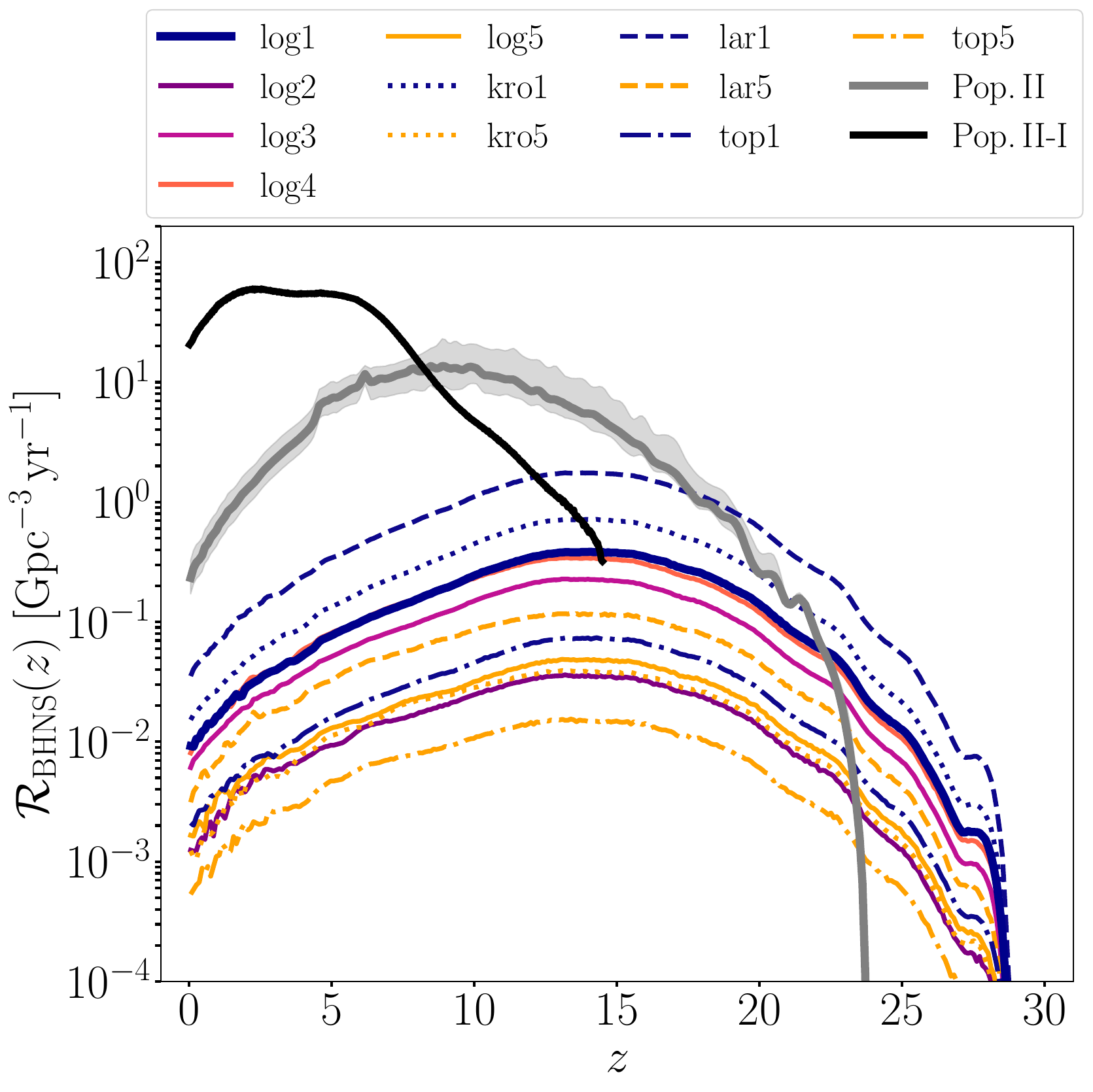}
    \caption{Merger rate density of BHNSs as a function of the redshift, $z$. The colored lines show the distributions for all the simulated Pop.~III models. The thick solid blue line represents our fiducial model for Pop.~III stars (log1). The thick solid gray line is the fiducial model for Pop.~II stars (kro1); the shaded gray area encompasses the uncertainties on galactic outflows and IMF parameters (see Sec.~\ref{sec:mrd_methods}). Finally, the thick solid black line represents the MRD of Pop.~II-I BHNSs estimated by \cite{sevn2023}.}
    \label{fig:mrd_bhns}
\end{figure}

Figure~\ref{fig:mrd_bhns} shows the MRD of BHNSs. Here, we consider all the models summarized in Table~\ref{table:ic_costa} for Pop.~III stars. For Pop.~II stars, instead, we take into account only the fiducial model (kro1). As a comparison, we also show the total MRD of Pop.~II-I BHNSs reported by \cite{sevn2023} (see Appendix~\ref{app:sevn} for details).

The MRD of BHNSs from Pop.~III stars peaks at $z\sim13$, where the assumed models span about three orders of magnitude, $\left[\mathcal{R}_{\rm BHNS}(z\sim13)\sim10^{-2}-2\,\rm Gpc^{-3}\,yr^{-1}\right]$. The position of the peak is the result of a convolution between the chosen Pop.~III star formation rate density \citep{hartwig2022} and the predominantly short delay times of BHNS mergers (see Sec.~\ref{sec:bhns_deltimes} and Fig.~\ref{fig:tdel_bhns}). At the peak, the largest MRD is associated with model lar1 (see Table~\ref{table:ic_costa} and Appendix~\ref{ap:isolated}). Model lar1 produces $\mathcal{R}_{\rm BHNS}(z\sim13)\sim1.7\,\rm Gpc^{-3}\,yr^{-1}$, which is approximately four times larger than the MRD peak of our fiducial model ($\mathcal{R}_{\rm BHNS}(z\sim13)\sim0.4\,\rm Gpc^{-3}\,yr^{-1}$). In fact, although top-heavy IMFs tend to produce more BH and neutron star progenitors, the expansion of these massive progenitors during their evolution very often leads to premature mergers of the stellar components,
inhibiting the formation of a compact binary (see Fig.~\ref{fig:r_evol}). At $z=0$, all the Pop.~III star models produce MRDs between $\sim10^{-3}$ and $\sim10^{-1}\,\rm Gpc^{-3}\,yr^{-1}$. In agreement with \cite{santoliquido2023}, we find that models with initial orbital periods from \cite{sana2012} produce larger MRDs than models relying on \cite{sb2013} (log2, log5, kro5, lar5, top5). In fact, the former have smaller orbital periods than the latter, favoring BHNS mergers. In Appendix~\ref{app:sfrd_comparison}, we repeat this analysis using different models of Pop.~III star formation rate densities.

The MRD of Pop.~II BHNSs peaks at $z\sim9$, reaching $\mathcal{R}_{\rm BHNS}(z\sim9)\sim14\,\rm Gpc^{-3}\,yr^{-1}$. At $z = 0$, Pop.~II BHNSs yield a MRD approximately seven times higher than that of Pop.~III stars under model lar1. The MRD from Pop.~II BHNSs continues to dominate over that of Pop.~III stars up to $z \sim 17$, even under the most favorable assumptions for Pop.~III stars.

Finally, Pop.~II-I BHNSs yield a MRD peaking at $z\sim2.5$ ($\mathcal{R}_{\rm BHNS}(z\sim2.5)\sim60\,\rm Gpc^{-3}\,yr^{-1}$) due to the combined effect of the star formation rate density \citep{madau2017} and the dependence on metallicity \citep{sevn2023}. Our comparison sample from \citet{sevn2023} including Pop.~I stars dominates over the contribution of Pop.~II stars up to $z\sim7$. 

Here, we note the seeming contradiction that the BHNS MRD from Pop.~II-I stars is lower than the BHNS MRD from Pop.~II stars alone at $z\sim{7-15}$. However, we note that the compact object MRD from Pop.~II-I stars reported by \citet{sevn2023} was obtained with several different assumptions compared to the new MRDs presented here. The main differences are the maximum ZAMS mass (assumed to be only 150 M$_\odot$ by \citealt{sevn2023}), and the different cosmic star formation rate density, $\psi{}(z,Z)$. Specifically, \cite{sevn2023} obtain $\psi{}(z,Z)$ from \cite{madau2017} assuming a log-normal metallicity distribution with a narrow dispersion $\sigma_\mathrm{Z}=0.1$ (see Fig.~\ref{fig:sfrd} and Appendix~\ref{app:sevn}), whereas here we assume self-consistent values of $\psi{}(z,Z)$ for both Pop.~II and III stars directly derived from \textsc{a-sloth} (\citealp{hartwig2022, hartwig2024}; Liu at al., in prep.). Assuming a narrow metallicity dispersion suppresses pockets of metal-poor star formation, as has already been discussed by \citet{sgalletta2025}. Table~\ref{table:mrd_bhns} reports a summary of the main results of this section.

\begin{table*}[ht!]
\centering
\caption{Merger rate density of BHNSs from Pop.~III and Pop.~II stars}
\begin{tabular}{c c c c c} 
 \hline
 Model & Population & $\mathcal{R}_{\rm BHNS}(z=0)$ & $\mathcal{R}_{\rm BHNS}(z=13.2)$ & $\mathcal{R}_{\rm BHNS}(z=8.9)$  \\ 
  &  & [$\rm Gpc^{-3}\,yr^{-1}$] & [$\rm Gpc^{-3}\,yr^{-1}$]  & [$\rm Gpc^{-3}\,yr^{-1}$] \\[0.5ex] 
 \hline\hline
 lar1 & Pop.~III & $3.5\times10^{-2}$ & 1.7 & 0.8 \\ 
 kro1 & Pop.~II & 0.2 & 6.5 & 13.7 \\
 \hline
\end{tabular}
\tablefoot{Column 1: Name of the model producing the largest MRD. In the case of Pop.~II stars, only the fiducial model kro1 was considered. Column 2: Stellar population. Column 3: MRD at $z=0$. Column 4: MRD at the peak of the Pop.~III distribution. Column 5: MRD at the peak of the Pop.~II distribution.}
\label{table:mrd_bhns}
\end{table*}

\begin{table*}[ht!]
\centering
\caption{Formation channels of BHNS mergers}
\begin{tabular}{c c c c c c c c c} 
 \hline
 Model & Population & $\eta_{\rm BHNS}$ & Ch. 0 & Ch. 1 & Ch. 2 & Ch. 3 & Ch. 4 & Ch. 5 \\ 
  &  & [$\msun^{-1}$] & (\%)  & (\%)  & (\%)  & (\%)  & (\%)  & (\%)\\[0.5ex] 
 \hline\hline
 log1 & Pop.~III & $1.2\times10^{-5}$ & 0.6 & 7.0 & 55.3 & 29.0 & 0 & 8.0\\ 
 kro1 & Pop.~II & $3.4\times10^{-5}$ & 0 & 44.8 & 46.9 & 1.4 & 0.2 & 6.7\\
 \hline
\end{tabular}
\tablefoot{Column 1: Name of the model. Column 2: Stellar population. Column 3: Merger efficiency of BHNSs, $\eta_{\rm BHNS}= N_{\rm BHNS,m}/M_{*}$, with $N_{\rm BHNS,m}$ the number of BHNS mergers. Column 4-9: Percentage of BHNS mergers happening through channel 0-5.}
\label{table:form_chan_bhns}
\end{table*}

\subsubsection{Delay times and formation channels} 
\label{sec:bhns_deltimes}

\begin{figure}[ht]
    \centering
    \includegraphics[width=0.9\columnwidth]{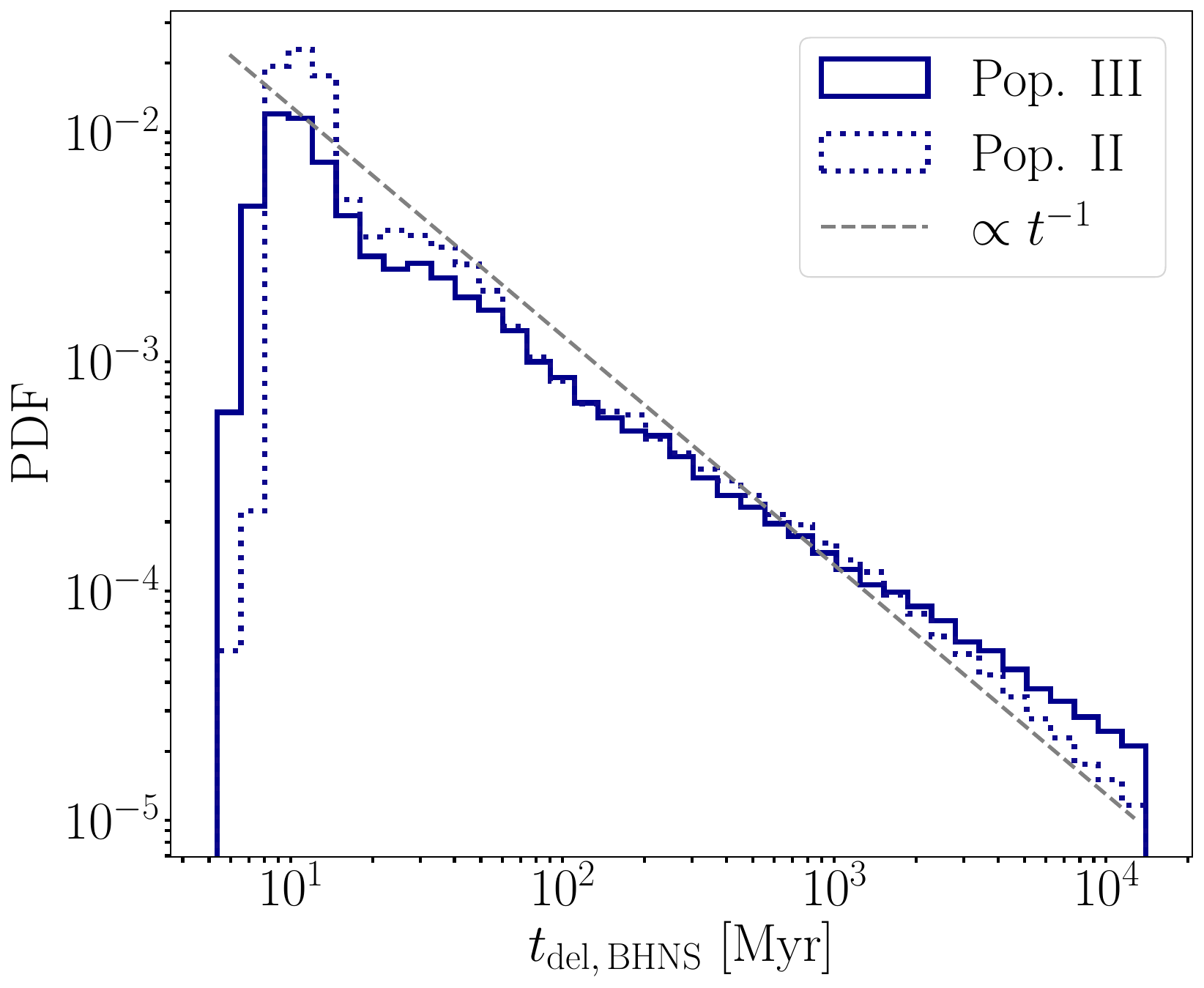}
    \caption{Distribution of delay times of BHNS mergers for the fiducial models of Pop.~III (continuous blue line) and Pop.~II stars (dotted blue line). The dashed gray line shows the $\propto t^{-1}$ trend from \cite{dominik2012}.}
     \label{fig:tdel_bhns}
\end{figure}

Figure~\ref{fig:tdel_bhns} shows the delay time (i.e., the time between the formation of the stellar binary system and its merger) for BHNS mergers in our fiducial Pop.~III and Pop.~II models. We compare these distributions with the $\propto t^{-1}$ trend from \cite{dominik2012}. This is the expected scaling of $t_{\rm del}$ if we consider orbital periods with a nearly log-uniform distribution (as in kro1 and log1, see Table~\ref{table:ic_costa}). Deviations from this trend suggest a final semimajor axis distribution that is significantly different from the initial one \citep{peters1964, sana2012}. In general, we find a good agreement between our simulations and the $\propto t^{-1}$ trend. 

We point out a deviation from the trend for Pop.~III BHNS mergers with $t_{\rm del}>10^3\,\rm Myr$, which is due to a majority of systems with semimajor axes larger than $\sim10\,\rm R_{\odot}$. In Table~\ref{table:form_chan_bhns}, we report the fraction of BHNS mergers happening through the formation channels previously described in Sec.~\ref{sec:sevn_form_chan}, and summarized in Table~\ref{table:descr_form_chan}. We find that Pop.~III BHNSs predominantly form via channel 2, which involves only stable mass transfer episodes and, as a result, does not significantly reduce the semimajor axis. The second most common pathway is channel 3, which is particularly relevant for systems with wide initial separations.

By contrast, Pop.~II BHNS mergers show an excess at short delay times ($t_{\rm del}\sim 15\,\rm Myr$). This is due to the large fraction ($\sim{45}\%)$ of Pop.~II BHNSs forming via channel 1, which shrinks semimajor axes effectively via the common envelope evolution, whereas channel 2 and 3 tend to have longer timescales.

\subsubsection{Primary mass and mass ratio} 
\label{sec:m1_q_bhns}

\begin{figure}[ht]
    \centering
    \includegraphics[width=0.9\columnwidth]{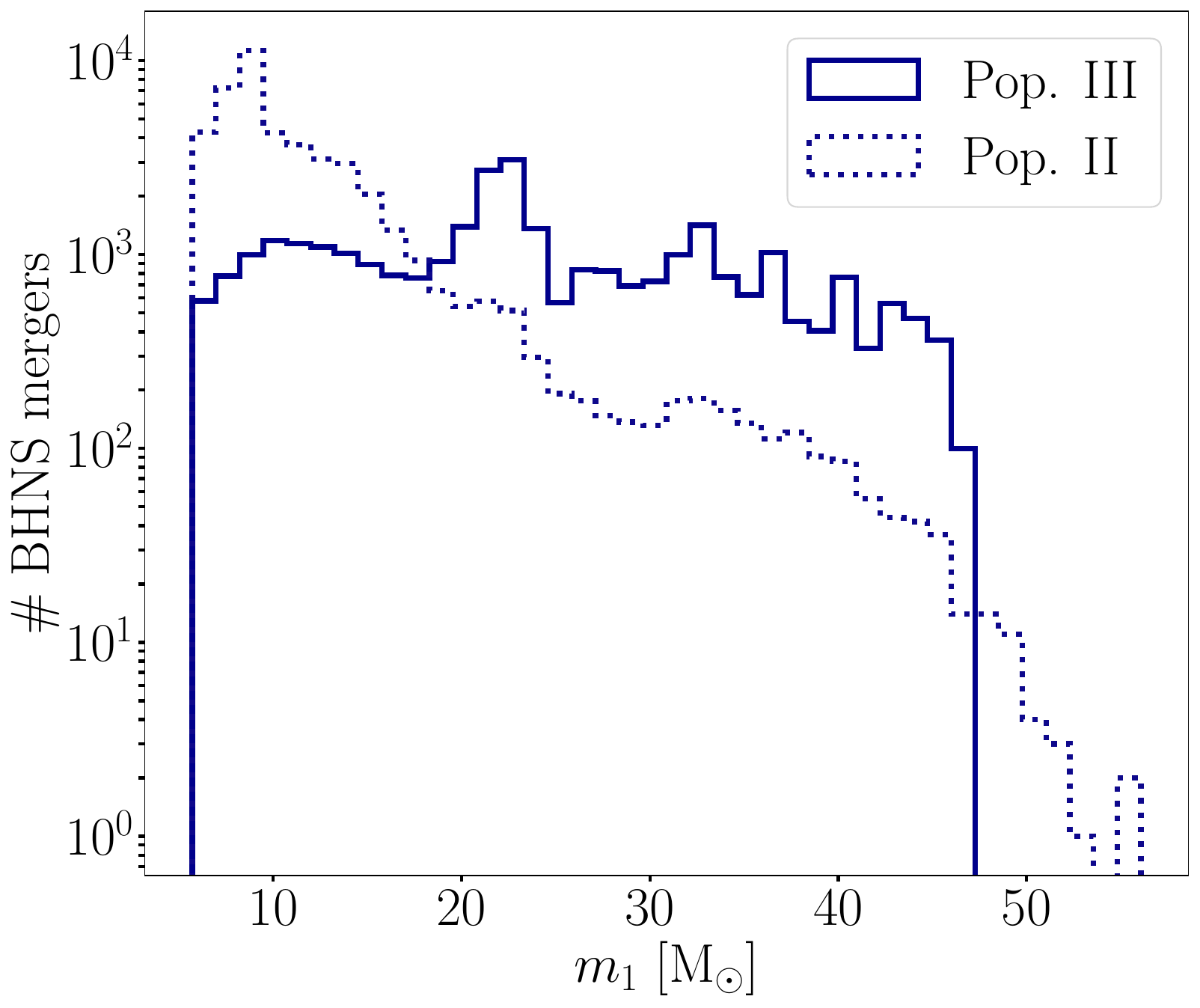}
    \caption{Distribution of primary mass of BHNS mergers for the fiducial models of Pop.~III (continuous blue line) and Pop.~II (dotted blue line).}
    \label{fig:m1_bhns}
\end{figure}

Figure~\ref{fig:m1_bhns} shows the primary mass (i.e., the BH mass) distribution of BHNS mergers. Pop.~II BHNS mergers show a peak at $m_1\lesssim15\,\msun$ \citep{giacobbo2018, floor2021}. After this peak, the number of BHNS mergers decreases with $m_1$ up to $\sim60\,\msun$. The BHNS mergers from Pop.~III stars, instead, have a flatter distribution in $m_1$ between 5 and $\sim47\,\msun$ and show a peak around $22\,\msun$. The different trends of these distributions are related to the dependence of the mass-radius relation on the metallicity, $Z$, of the stars. In fact, regardless of the formation channel of BHNS mergers, Pop.~II star binaries with $m_{1,\rm ZAMS}>40\,\msun$ will tend to prematurely merge unless $a_{\rm ZAMS}\gtrsim10^3\,\rm R_{\odot}$. Population III star binaries with high $m_{1,\rm ZAMS}$ will instead expand less during their evolution, and avoid premature mergers if $a_{\rm ZAMS}\gtrsim400\,\rm R_{\odot}$. As is reported in Table~\ref{table:form_chan_bhns}, the lack of a prominent peak at $m_1\leq20\,\msun$ and the choice of a top-heavier IMF lead to a lower merger efficiency for Pop.~III BHNSs, $\eta_{\rm BHNS}$. 

\begin{figure}[ht]
    \centering
    \includegraphics[width=\columnwidth]{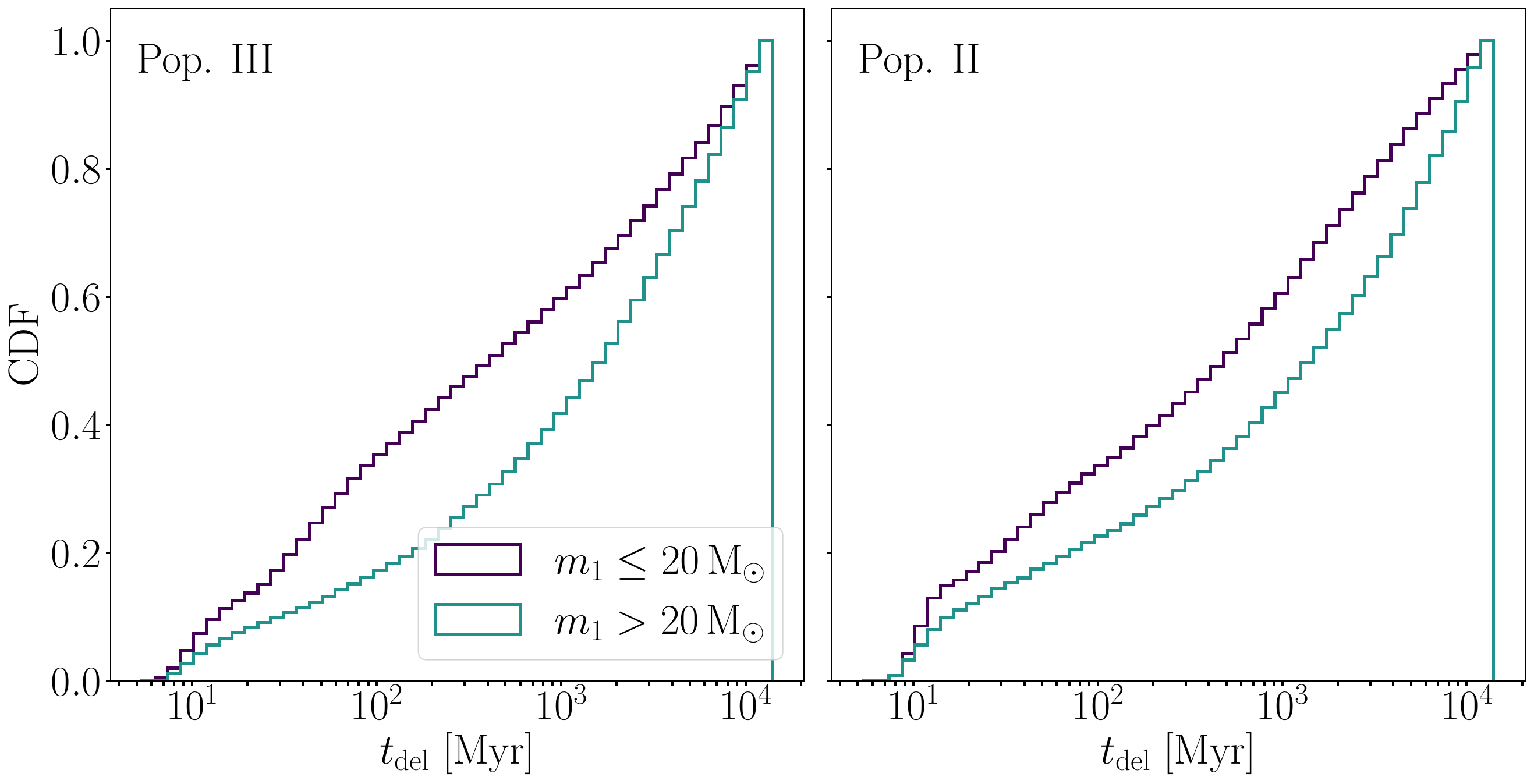}
    \caption{Cumulative distribution of the delay times, $t_{\rm del}$, of BHNS mergers in two ranges of primary mass, $m_1$. We represent in violet mergers with $m_1\leq20\,\msun$ and in blue-green mergers with $m_1 > 20\,\msun$. We show the results for the fiducial models of Pop.~III (left), and Pop.~II progenitors (right).}
    \label{fig:tdel_m1_bhns}
\end{figure}

Figure~\ref{fig:tdel_m1_bhns} shows the cumulative distribution of delay times for two primary mass bins: light ($m_1\leq 20\,\msun$) and heavy ($m_1>20\,\msun$) primary BHs. High-mass Pop.~III BHNS mergers are associated with longer delay times ($t_{\rm del}>10^3\,\rm Myr$) compared to light systems. This means that, at low redshifts, we might be able to observe GW signals from BHNSs with $m_1>20\,\msun$ deriving from the evolution of metal-free binaries. In fact, more than $85\%$ of Pop.~III BHNS mergers have $m_1>20\,\msun$, while the fraction reduces to $\sim11\%$ for Pop.~II. As a consequence, even though at low redshifts the MRD of Pop.~II BHNSs is about one order of magnitude larger than the one of Pop.~III BHNSs (see Fig.~\ref{fig:mrd_bhns}), we have eight times more mergers from Pop.~III BHNSs with $m_1>20\,\msun$ compared to Pop.~II. The contribution of Pop.~III BHNS mergers to the currently observed GW events is discussed further in Sec.~\ref{sec:gw_comparison}. 

The smaller population of Pop.~II BHNS mergers with $m_1> 20\,\msun$ also tends to occur at long delay times, following a trend similar to Pop.~III (see bottom panel of Fig.~\ref{fig:tdel_m1_bhns}). In both cases, in fact, binaries with massive $m_{\rm ZAMS, 1}$ avoid premature mergers only when their initial separations are wide, leading to longer $t_{\rm del}$.

The delay times of light BHNS mergers, instead, have different trends for the two stellar populations. The subdominant Pop.~III BHNS mergers with $m_1\leq20\,\msun$ have smaller delay times with respect to Pop.~II BHNSs, owing to the greater compactness of Pop.~III progenitors at low $m_{\rm 1,ZAMS}$ (see Fig.~\ref{fig:r_evol}), which facilitates the formation of close compact binaries.

\begin{figure}[ht]
    \centering
    \includegraphics[width=0.9\columnwidth]{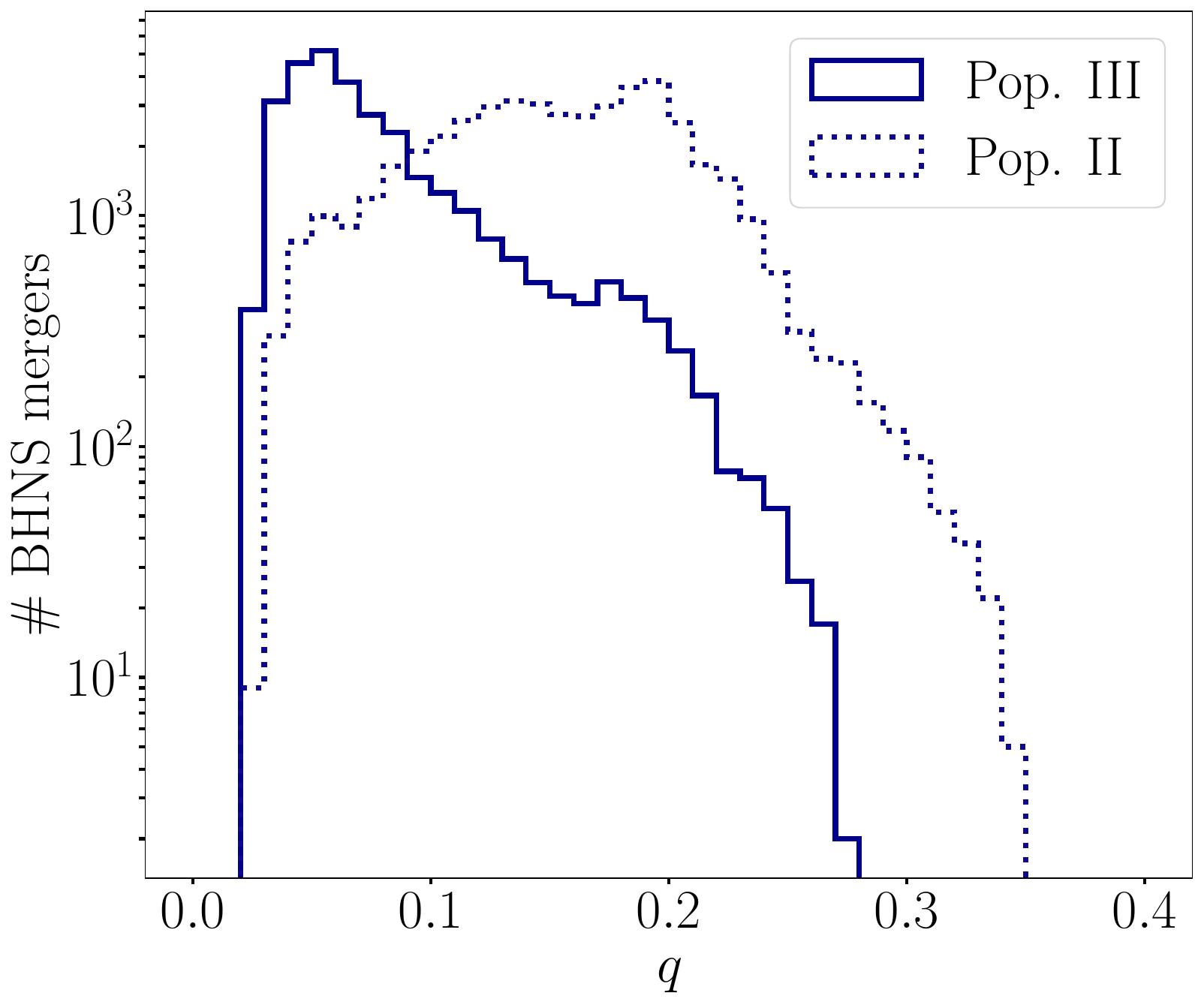}
    \caption{Distribution of mass ratios, $q$, of BHNS mergers for the fiducial models of Pop.~III (continuous blue line) and Pop.~II (dotted blue line).}
     \label{fig:q_bhns}
\end{figure}

Finally, Fig.~\ref{fig:q_bhns} shows the distributions of the mass ratio, $q$, of the simulated BHNS mergers. Pop.~III stars yield a distribution that peaks at $q<0.1$ and extends up to $q\sim0.3$; in fact, as we have seen, metal-free stars favor BHNS mergers with a large primary mass, resulting in small values of $q$. In contrast, Pop.~II BHNSs merge with a mass ratio that peaks at $q\sim0.2$ and extends to larger values, up to $q\sim0.35$. In accordance with both \cite{floor2021} and \cite{giacobbo2018}, this is a consequence of the peak at $m_1<15\,\msun$ (Fig.~\ref{fig:m1_bhns}) and of the lower mass gap, between $2$ and $5\,\msun$. These distinguishing features are particularly interesting because they can also contribute to a distinction between GW events from Pop.~III and Pop.~II binaries (see Sec.~\ref{sec:gw_comparison}).

\subsection{Binary neutron stars}

In this section, we present the properties of BNS mergers from Pop.~III and Pop.~II stars. Firstly, we display the MRD of BNSs from different simulated models of Pop.~III stars and compare them to the MRD of Pop.~II BNSs. We then discuss the delay time distributions. 

\subsubsection{Merger rate density} 
\label{sec:mrd_bns}

\begin{figure}[ht]
    \centering
    \includegraphics[width=0.9\columnwidth]{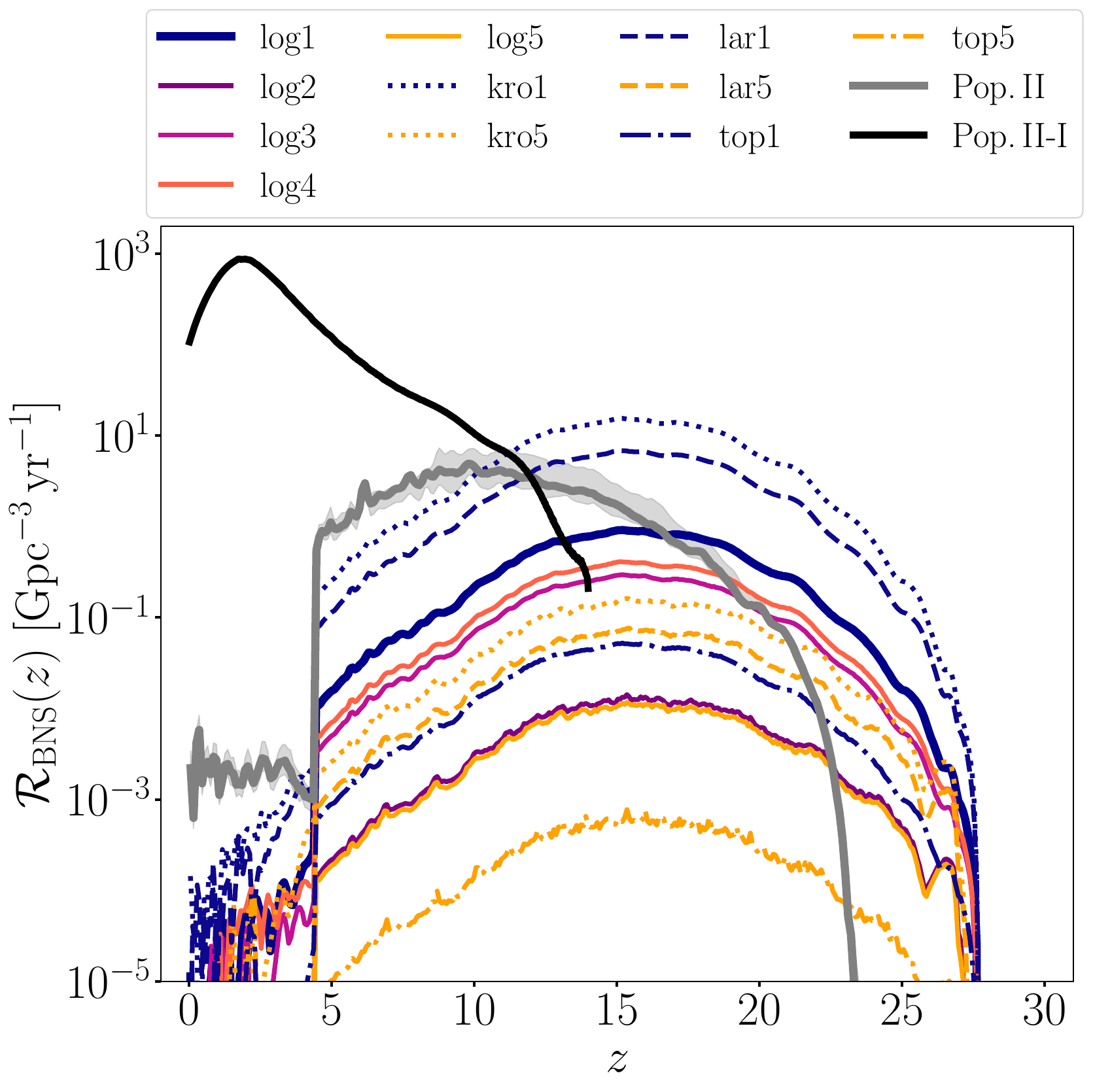}
    \caption{Same as Fig.~\ref{fig:mrd_bhns} but for BNS mergers.}
    \label{fig:mrd_bns}
\end{figure}

Figure~\ref{fig:mrd_bns} shows the MRD of BNSs from Pop.~III and Pop.~II stars. For Pop.~III stars, we take into account all the simulated models reported in Table~\ref{table:ic_costa}. In the case of Pop.~II stars, instead, we only consider the fiducial model kro1. As we did for BHNSs, we compare these MRDs with the one derived by \citet{sevn2023}, which includes higher-metallicity stars.

The MRD of BNSs from Pop.~III stars peaks at $z\sim15$, for all considered models. Here, the different models we simulated span five orders of magnitude: $\left[\mathcal{R}_{\rm BNS}(z\sim15)\sim10^{-3}-15\,\rm Gpc^{-3}\,yr^{-1}\right]$. Because of the extremely short BNS delay times (see Sec.~\ref{sec:bns_deltimes} and Fig.~\ref{fig:tdel_bns}), the MRDs from Pop.~III stars peak in the vicinity of the peak of the chosen star formation rate density. At $z < 5$, their behavior is driven by the convolution of the short delay times ($<50\,\mathrm{Myr}$) with the chosen $\psi(z',Z)$ (see Fig.~\ref{fig:sfrd}). Since Pop.~III BNSs lack systems with long delay times, their MRDs rapidly decline once the underlying star formation rate density drops, without extending significantly beyond the star formation rate density cutoff.

Model kro1 produces the highest MRD for Pop.~III stars, with $\mathcal{R}_{\rm BNS}(z)\sim15\,\rm Gpc^{-3}\,yr^{-1}$ at $z\sim15$. The peak of our fiducial Pop.~III model (log1) is an order of magnitude lower. In fact, kro1 produces less massive BNS progenitors involved in premature mergers. As is discussed in Sec.~\ref{sec:mrd_bhns}, models adopting the initial orbital period distribution from \cite{sb2013} also result in lower MRDs in this case. In Appendix~\ref{app:sfrd_comparison}, we take into account the impact of Pop.~III star formation rate density models on the MRDs.

The BNS MRD from Pop.~II stars dominates over the one from Pop.~III stars up to $z\sim10$, where it reaches a value of $\mathcal{R}_{\rm BNS}(z\sim10) \sim 5\,\rm Gpc^{-3}\,yr^{-1}$. There, the MRD of Pop.~III BNSs from model kro1 is within the same order of magnitude. As was pointed out for Pop.~III BNSs, Pop.~II BNS mergers also have extremely short delay times, leading to a drop in the MRD at $z<5$. 

Figure~\ref{fig:r_evol} shows that Pop.~III stars with $m_{\rm ZAMS}\lesssim20\,\msun$ are more compact with respect to Pop.~II stars. As a consequence, they produce, on the one hand, less premature mergers and, on the other, a higher BNS merger efficiency ($\eta_{\rm BNS}\sim10^{-4}\,\msun^{-1}$ for model kro1). However, as is shown in Table~\ref{table:mrd_bns}, convolving the merger efficiency with the star formation rate density models yields a MRD from Pop.~II BNSs that exceeds that of Pop.~III by at least one order of magnitude at $z<10$. 

For comparison, we show that the MRD of BNSs born from Pop.~II-I progenitors \citep{sevn2023} peaks at $z\sim 2$ ($\mathcal{R}_{\rm BNS}(z\sim2)\sim10^3\,\rm Gpc^{-3}\,yr^{-1}$), consistently with the peak of the star formation rate density from \cite{madau2017}. The contribution of BNSs born from more metal-rich progenitors dominates the MRDs below $z\sim11$.

\begin{table*}[ht!]
\centering
\caption{Merger rate density of BNSs from Pop.~III and Pop.~II stars.}
\begin{tabular}{c c c c c} 
 \hline
 Model & Population & $\mathcal{R}_{\rm BNS}(z=0)$ & $\mathcal{R}_{\rm BNS}(z=15.2)$ & $\mathcal{R}_{\rm BNS}(z=9.8)$  \\ 
  &  & [$\rm Gpc^{-3}\,yr^{-1}$] & [$\rm Gpc^{-3}\,yr^{-1}$]  & [$\rm Gpc^{-3}\,yr^{-1}$] \\[0.5ex] 
 \hline\hline
 kro1 & Pop.~III & $1.4\times10^{-4}$ & 15.4 & 3.2 \\ 
 kro1 & Pop.~II & $2.2\times10^{-3}$ & 1.6 & 4.9 \\
 \hline
\end{tabular}
\tablefoot{Same as Table~\ref{table:mrd_bhns} but for the MRD of BNSs from Pop.~III and Pop.~II stars.}
\label{table:mrd_bns}
\end{table*}

\begin{table*}[ht!]
\centering
\caption{Formation channels of BNS mergers.}
\begin{tabular}{c c c c c c c c c} 
 \hline
 Model & Population & $\eta_{\rm BNS}$ & Ch. 0 & Ch. 1 & Ch. 2 & Ch. 3 & Ch. 4 & Ch. 5 \\ 
  &  & [$\msun^{-1}$] & (\%)  & (\%)  & (\%)  & (\%)  & (\%)  & (\%)\\[0.5ex] 
 \hline\hline
 log1 & Pop.~III & $6.2\times10^{-6}$ & 0 & 87.1 & 0 & 3.6 & 0.2 & 9.1\\ 
 kro1 & Pop.~II & $4.3\times10^{-6}$ & 0 & 45.4 & 0 & 53.5 & 0.4 & 0.7\\
 \hline
\end{tabular}
\tablefoot{Same as Table~\ref{table:form_chan_bhns} but for BNS mergers.}
\label{table:form_chan_bns}
\end{table*}

\subsubsection{Delay times and formation channels} 
\label{sec:bns_deltimes}

\begin{figure}[ht]
    \centering
    \includegraphics[width=0.9\columnwidth]{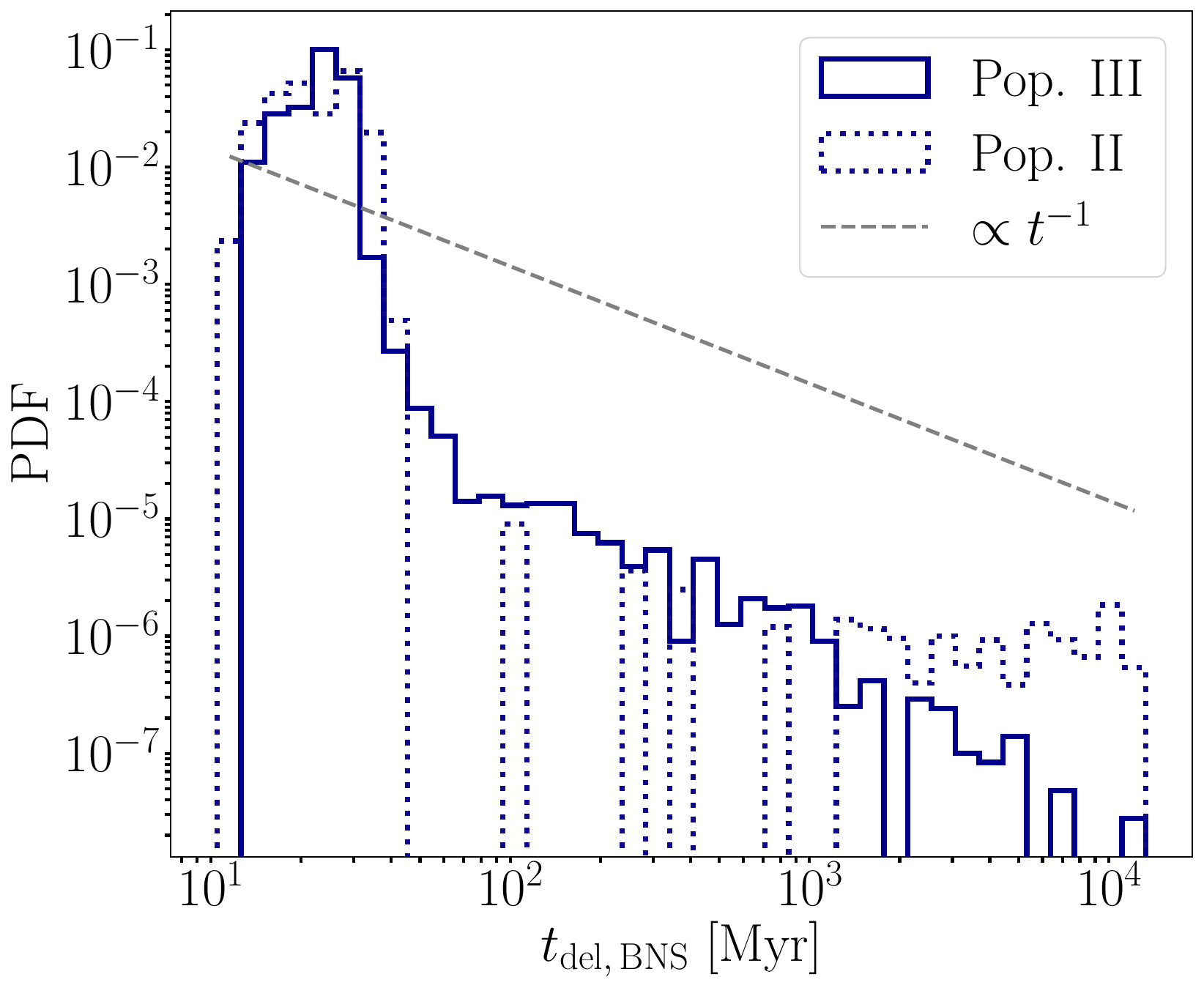}
    \caption{Same as Fig.~\ref{fig:tdel_bhns} but for BNS mergers.}
     \label{fig:tdel_bns}
\end{figure}

Figure~\ref{fig:tdel_bns} shows the distribution of delay times of BNS mergers for the fiducial models of Pop.~III and Pop.~II stars. Both populations result in distributions deviating significantly from the $\propto t^{-1}$ trend predicted by \cite{dominik2012}. The delay times of BNS mergers show a peak at $\sim20\,\rm Myr$, and a common trend up to $t_{\rm del}\sim10^3\,\rm Myr$, after which we observe an excess of BNS mergers from Pop.~II stars. 

The peak at $\sim 20\,\rm Myr$ is the result of very efficient orbital shrinking after one (or more) common envelope event(s). As a matter of fact, Table~\ref{table:form_chan_bns} shows that BNS mergers take place when a binary system undergoes at least one common envelope event during its evolution (channels 1, 3, 4, and 5). As a consequence, more than $98\,\%$ of the merging BNSs have $a<1\,\rm R_{\odot}$ at the time of formation of the second neutron star. 

The excess of Pop.~II systems with $t_{\rm del}>10^3\,\rm Myr$ can also be explained by looking at Table~\ref{table:form_chan_bns}. In fact, half of the BNS mergers from Pop.~II stars form through channel 3 (see Table~\ref{table:descr_form_chan}), which favors binaries with $a_{\rm ZAMS}\sim200-5\times10^4\,\rm R_{\odot}$. Indeed, Fig.~\ref{fig:r_evol} shows that Pop.~II stars with $m_{\rm ZAMS}\sim 8-20\,\msun$ tend to expand to larger radii with respect to Pop.~III stars in the same mass range. Consequently, only binary stars with large initial separations will avoid to merge prematurely. We find that a fraction of the binaries evolving through channel 3 yields BNSs with $a>2\,\rm R_{\odot}$, which will need $\geq 10^3\,\rm Myr$ to merge.

Lastly, Fig.~\ref{fig:tdel_bns} points out the lower merger efficiency of Pop.~II BNSs, with an evident lack of systems merging with $50<t_{\rm del}<10^3\,\rm Myr$. In Table~\ref{table:form_chan_bns} shows that, for the chosen fiducial models, the merger efficiency of Pop.~II BNSs is $\sim 1.4$ times lower than the one of Pop.~III BNSs.

\section{Discussion} 
\label{discussion}

In the following, we discuss the impact of the initial conditions on the properties of BHNS mergers. We then compute the detectability-conditioned Bayes factors for three GW events comparing a Pop.~III origin to a Pop.~II one.

\subsection{Impact of initial conditions on distributions of $m_1$ and $q$} \label{sec: bhns_primary_models}

\begin{figure*}[ht]
    \centering
    \includegraphics[width=0.9\textwidth]{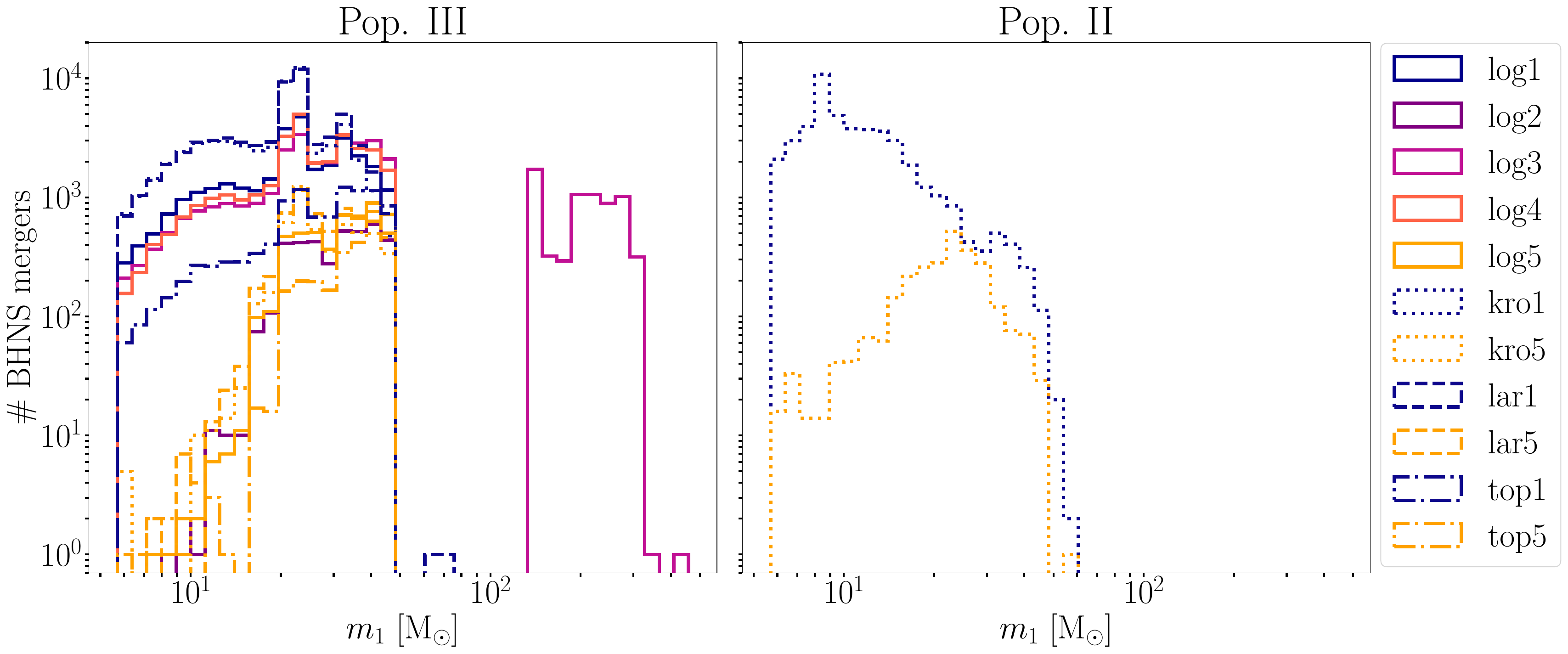}
    \caption{Primary BH mass distribution of BHNS mergers. The left-hand panel shows the distributions for all the simulated models of Pop.~III stars reported in Table~\ref{table:ic_costa}. The right-hand panel shows the $m_1$ distributions of Pop.~II stars for models kro1 and kro5.}
     \label{fig:bhns_m1_all}
\end{figure*}

Figure~\ref{fig:bhns_m1_all} shows the distribution of the primary mass, $m_1$, of BHNS mergers for all the simulated models of Pop.~III stars (Table~\ref{table:ic_costa}). As a comparison, we also show the distributions for the two Pop.~II star models assuming a Kroupa IMF (kro1, kro5; \citealp{kroupa2001}). 

All Pop.~III star models produce BHNS mergers with a primary mass of between $\sim6\,\msun$ and $\sim50\,\msun$, with the exception of log3, which pushes $m_1$ above the upper mass gap, up to $\sim 400\,\msun$. In fact, the initial mass-ratio distribution adopted in model log3 (“sorted;” see Table~\ref{table:ic_costa}) favors the formation of binaries with \(m_{1,\rm ZAMS} \gtrsim 200\,\msun\) and \(m_{2,\rm ZAMS} \lesssim 25\,\msun\). These high-mass binaries typically have initial separations of \(a_{\rm ZAMS} > 10^3\,\rm R_{\odot}\) and merge predominantly through channel~3 (Table~\ref{table:descr_form_chan}), which favors unequal-mass systems \citep{costa2023}.

Models with initial orbital parameters from \cite{sb2013} (log2, log5, kro5, lar5, and top5) tend to produce BHNS mergers that have $m_1\gtrsim15\,\msun$ and peak at $m_1\gtrsim20\,\msun$. These models yield fewer BHNS mergers because of the wide initial semimajor axes; merging BHNSs have $a_{\rm ZAMS}\sim10^3-10^5\,\rm R_{\odot}$ and evolve mainly through channel 3. Unlike log3, these models rely on an initial mass-ratio distribution from \cite{sb2013} with $q_{\rm min}=0.1$, which favors the formation of BHNSs with large $m_{\rm 1,ZAMS}$, but does not allow systems with $m_{\rm 1,ZAMS}>200\,\msun$.

Finally, models based on \cite{sana2012} yield a flatter $m_1$ distribution, with a sharp peak around $m_1\sim22\,\msun$. We point out that model kro1 produces more mergers by a factor of three with respect to the Pop.~III fiducial model log1. 

Consequently, all the simulated Pop.~III star models show a nearly flat distribution of BH mass for $m_1>20\,\msun$. Moreover, in Sec.~\ref{sec:m1_q_bhns}, we have seen that these massive systems usually merge with large delay times. Thus, it is important to also take into account the contribution of Pop.~III massive BHNS mergers at low redshifts (see Sec.~\ref{sec:gw_comparison}). 

Pop.~II BHNS mergers from model kro1 show a peak at $m_1\sim10\,\msun$ and follow a decreasing trend up to $\sim60\,\msun$ (see Sec.~\ref{sec:m1_q_bhns}). As in the case of Pop.~III stars, Pop.~II BHNS mergers from model kro5 \citep{sb2013} mostly have $m_1>15\,\msun$. Model kro5 is thus able to reproduce massive BHNS mergers without the help of stellar dynamics \citep{rastello2020, mas2020, mas2024}. Nevertheless, because of their larger radial expansion with respect to Pop.~III stars, Pop.~II binaries are also able to produce more BHNS mergers with low-mass primaries. 

\begin{figure*}[ht]
    \centering
    \includegraphics[width=0.9\textwidth]{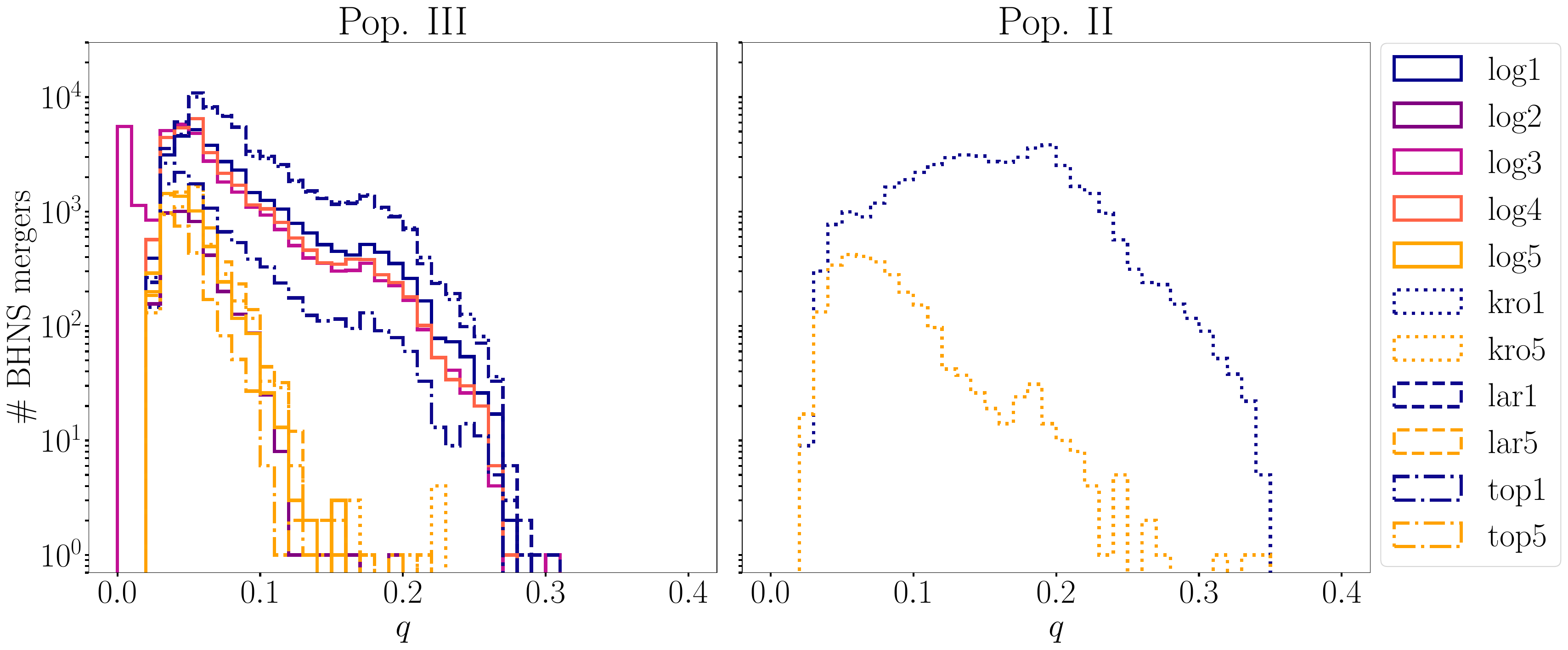}
    \caption{Same as Fig.~\ref{fig:bhns_m1_all} but for the mass ratio $q$.}
     \label{fig:bhns_q_all}
\end{figure*}

Figure~\ref{fig:bhns_q_all} shows the mass-ratio $q$ distributions for all the simulated models. Starting from Pop.~III stars, we notice that models with initial orbital periods from \cite{sana2012} peak at $q\sim0.1$ and extend up to $q\sim0.3$. Models based on \cite{sb2013}, instead, get to a maximum of $q\sim0.2$. We point out again model log3, producing BHNS mergers with extremely low mass ratios; this feature was already found and discussed by \cite{costa2023} in the case of binary BHs. Moving on to Pop.~II BHNS mergers, we see that model kro1 produces a peak at $q\sim0.2$ and extends to larger mass ratios with respect to its metal-free counterpart (see also Sec.~\ref{sec:m1_q_bhns}). Due to its primary mass distribution, model kro5 shows intermediate features between Pop.~III BHNS mergers with initial parameter distributions from \cite{sb2013} and from \cite{sana2012}.

\subsection{Model selection with GW events} 
\label{sec:gw_comparison}
Here, we compare our models by calculating the detectability-conditioned posterior odds for GW events with BHNS components. To carry out this comparison, we followed the formalism presented in \cite{mould2023}, which we briefly summarize in Appendix~\ref{ap:gw_comparison}. Selection effects were computed using public LVK injections \citep{KAGRA:2023pio} with a detection threshold in the false-alarm rate of $<0.25\,\mathrm{yr}^{-1}$ \citep{abbottO3popandrate}.

Table~\ref{table:odds} presents the detectability-conditioned Bayes factors, \(\ln \mathcal{D}_{A/B}\), where model \(A\) corresponds to a Pop.~III model (as is listed in Table~\ref{table:ic_costa}) and model \(B\) represents the fiducial Pop.~II model (kro1). These values were obtained by convolving the merger catalogs generated with \textsc{sevn} with \textsc{cosmo}$\mathcal{R}$\textsc{ate}, following the methodology described in Sec.~\ref{sec:mrd_methods}, and considering only mergers occurring at redshifts \(z < 2\). We computed Bayes factors for three GW events: GW190917, GW191219, and GW200105 \citep{gw200105_gw200115,gwtc2.1,gwtc3_2023}. Our analysis includes GW190917, with a secondary mass within the lower mass gap (\(2-5\,\msun\)). The waveform parameters used to evaluate the Bayes factors are the primary mass, mass ratio, and redshift of each event. In this analysis, we only consider events with sufficient overlap between the observed parameters and the simulated distributions to ensure meaningful Bayes factor estimates. A decisive preference between models is typically indicated when \(|\ln \mathcal{D}_{A/B}| \gtrsim 5\) \citep{jeffreys1939}.

Among the selected GW events, only one (GW191219) has a primary mass of \(m_1 > 20\,\msun\). It is generally expected that Pop.~II BHNS mergers with such high primary masses form in star clusters \citep{rastello2020,mas2020,mas2024}. However, as is illustrated in Figs.~\ref{fig:m1_bhns} and \ref{fig:tdel_m1_bhns}, an alternative formation channel could derive from the evolution of isolated Pop.~III binaries.

GW191219 favors a Pop.~III star origin. This system is characterized by a highly asymmetric mass ratio (\(q \sim 0.04\)), leading to higher Bayes factors for models that favor mergers with extremely low mass ratios, such as kro5, lar5, log3, log4, log5, and top5. This outcome is consistent with Fig.~\ref{fig:bhns_q_all}, which shows that all simulated Pop.~III models predominantly yield BHNS mergers with \(q < 0.1\). It should be noted that the false alarm rate associated with GW191219 is $\sim 4\,\rm{yr}^{-1}$, which is not consistent with the threshold typically adopted for estimating selection effects in the calculation of detection-conditioned Bayes factors. Nevertheless, the detectability-free Bayes factors, $\mathcal{B}_{\rm Pop.~III/Pop.~II}$, for this event are found to be in close agreement with the detectability-weighted values reported in the table. The remaining two GW events, GW190917 and GW200105, both have primary masses of \(m_1 < 20\,\msun\) and show stronger support for a Pop.~II formation channel compared to the Pop.~III scenario. 

As a consequence, we find that Pop.~III BHNS mergers contribute to the population of systems with a high primary mass and a low mass ratio at low redshift more than Pop.~II stars. In contrast, for $m_1 < 15\,\msun$, Pop.~II binaries remain the dominant formation channel for the low-redshift BHNS mergers. 

\begin{table*}[ht!]
\centering
\caption{Detectability-conditioned Bayes factors, $\ln \mathcal{D}$, for Pop.~III models against the fiducial Pop.~II model}
\renewcommand{\arraystretch}{1.3} 
\resizebox{\textwidth}{!}{
\begin{tabular}{c c c | c c c c c c c c c c c} 
 \hline
 \multicolumn{3}{c|}{GW signals} & \multicolumn{11}{c}{$\ln\mathcal{D}_{\rm Pop.\,III/Pop.\,II}$} \\
 Name & $m_{1, \rm s}\,[\msun]$ & $m_{2, \rm s}\,[\msun]$ & kro1 & kro5 & lar1 & lar5 & log1 & log2 & log3 & log4 & log5 & top1 & top5 \\ 
 \hline\hline
 GW190917 & $9.7_{-3.9}^{+3.4}$ & $2.1^{+1.1}_{-0.4}$ 
         & $-1.1$ & $-3.3$ & $-0.4$ & $-4.2$ & $-1.7$ & $-3.9$ & $-1.3$ & $-1.6$ & $-4.2$ & $-1.1$ & $-83.2$ \\
 GW191219 & $31.1_{-2.8}^{+2.2}$ & $1.17_{-0.06}^{+0.07}$ 
          & 2.2 & 3.7 & 1.9 & 4.5 & 2.1 & 1 & 2.3 & 2.2 & 3.4 & 1.7 & 4.1 \\
 GW200105 & $9.1_{-1.7}^{+1.7}$ & $1.91_{-0.24}^{+0.33}$ 
           & $-4.4$ & $-11.2$ & $-4.7$ & $-9.9$ & $-2.7$ & $-16$ & $-3.7$ & $-3$ & $-14.8$ & $-3.3$ & $-512.4$ \\
 \hline
\end{tabular}
} 
\tablefoot{Detectability-conditioned Bayes factors, $\ln \mathcal{D}$, for Pop.~III models, evaluated against the fiducial Pop.~II model (kro1), for GW events classified as BHNS mergers or containing one component in the lower mass gap. The selected events were detected during the first three observing runs of LVK. Columns 1–3: Event name and source-frame component masses. Columns 4–14: $\ln \mathcal{D}$ values for each Pop.~III model relative to the Pop.~II reference model.}
\label{table:odds}
\end{table*}

\subsection{Caveats and future prospects}
As is shown in Table~\ref{table:ic_costa}, we have explored a wide range of Pop.~III IMFs and orbital parameter models. However, our results are still affected by several uncertainties in stellar and binary evolution.

Stellar structure assumptions are a key source of uncertainty. For instance, suppressing core overshooting leads to more compact stars, which enhances the merger rate and increases the number of systems forming BHs above the mass gap \citep{tanikawa2021, tanikawa2021b, tanikawa2022, tanikawa2023, tanikawa2024}. Moreover, Pop.~III stars are expected to be rapid rotators, potentially undergoing chemically homogeneous evolution. As is shown by \cite{santoliquido2023} and \cite{mestichelli2024}, this channel produces very compact stars, again boosting merger rates and potentially leading to BHNS mergers with primary masses above the upper mass gap. 

Another uncertainty concerns the initial distributions of the orbital period, mass ratio, and eccentricity. Current hydrodynamical simulations are limited and typically biased toward longer orbital periods (e.g., \citealp{sb2013, sugimura2020, park2023, klessen2023}). To account for this, we adopt two prescriptions for orbital parameter distributions, from \citet{sana2012} and \citet{sb2013}, chosen to span a plausible range of outcomes. We also note that the distributions of \citet{sana2012}, derived for local O- and B-type stars, have already been employed in several recent studies \citep{tanikawa2022, tanikawa2023, costa2023, santoliquido2023, tanikawa2024, mestichelli2024}.

Supernova physics represents another major source of uncertainty. Throughout this work we have adopted the rapid SN model of \citet{fryer2012}. Alternative prescriptions could alter the component masses and mass-ratio distributions of compact-object binaries. For instance, the delayed SN model of \citet{fryer2012} allows neutron-star masses to extend up to $3\,\msun$ and BH masses down to the same value. This would shift the resulting mass-ratio distribution toward higher values compared to Fig.~\ref{fig:q_bhns}. We shall explore the impact of SN models on BNS and BHNS mergers from Pop.~III in future work. In addition, recent studies \citep{disberg2025a, disberg2025b} indicate that the natal kicks derived by \citet{hobbs2005} are likely overestimated by $\sim50\%$, due to an erroneous histogram representation. Correcting for this yields systematically lower kicks, which increases the survival probability of binaries containing neutron stars. Nevertheless, even the revised kick distributions remain sufficiently strong to disrupt a large fraction of systems, particularly those with wide orbits. Updated models for SN natal kicks will be taken into account as well for future work.

Finally, in this work we have focused exclusively on Pop.~III and early Pop.~II stars. This choice was motivated by our goal of investigating their role at high redshift, where they are expected to be most relevant, and in view of the future observations that will be conducted by third-generation GW interferometers. We have chosen to also keep this work focused on metal-poor populations in the case of the model selection performed in Sec.~\ref{sec:gw_comparison}.

\section{Summary} \label{summary}
We have investigated the properties of BHNS and BNS mergers from Pop.~III and Pop.~II stars, by means of binary population synthesis simulations performed with \textsc{sevn} \citep{costa2023}. These merger catalogs encompass the uncertainties on IMFs and orbital parameters of Pop.~III stars. We have estimated the merger rate densities (MRDs) of both BHNSs (Fig.~\ref{fig:mrd_bhns}) and BNSs (Fig.~\ref{fig:mrd_bns}) from Pop.~III stars and compared them to the MRDs of Pop.~II stars.

We find that the MRD of BHNSs from Pop.~III stars ranges from $\sim10^{-2}$ to $2\,\rm Gpc^{-3}\,yr^{-1}$ at redshift $z\sim{13}$. Models with initial orbital parameters from \cite{sb2013} produce smaller MRDs with respect to models with distributions from \cite{sana2012}. This behavior was already pointed out in \cite{santoliquido2023} in the case of binary BH mergers. The MRD of BHNSs from Pop.~II stars dominates over the one from Pop.~III stars at least up to $z\sim17$. The delay times, $t_{\rm del}$, of BHNS mergers for our fiducial models (Fig.~\ref{fig:tdel_bhns}) follow the $\propto t^{-1}$ trend from \cite{dominik2012} quite closely.

We compare the primary BH mass, $m_1$, distribution of BHNS mergers for the fiducial models, and find that BHs from Pop.~III stars tend to produce a flat distribution in $m_1$ with a peak around $22\,\msun$, while BHs from Pop.~II stars present a peak at $m_1<15\,\msun$ and a decreasing trend up to $\sim 60\,\msun$ (Fig.~\ref{fig:m1_bhns}). These different trends in $m_1$ result in different trends in the mass ratio, $q = m_2/m_1$ (Fig.~\ref{fig:q_bhns}). The BHNSs from Pop.~III stars yield a distribution peaking at $q<0.1$ and extending up to $q\sim0.3$, whereas Pop.~II BHNS mergers peak at $q\sim0.2$ and extend up to $q\sim0.35$. 

Population III high-mass BHNSs merge preferentially with large delay times, $t_{\rm del}>10^3\,\rm Myr$ (Fig.~\ref{fig:tdel_m1_bhns}). Consequently, second-generation interferometers might be able to detect BHNS mergers with $m_1>20\,\msun$ from a metal-free population in the local Universe. 

The BNSs from Pop.~III and~II stars mainly merge with short delay times ($t_{\rm del}<50\,\rm Myr$) because of their formation channels, which usually involve one or more common envelope events (Fig.~\ref{fig:tdel_bns}). As a result, the peaks of the MRDs of BNSs from both Pop.~III and Pop.~II stars are close to the peaks of their respective star formation rate density models \citep{hartwig2022, hartwig2024, liu2025}.
We find that the MRDs of metal-free BNSs range from $\sim10^{-3}$ to $\sim15\,\mathrm{Gpc}^{-3}\,\mathrm{yr}^{-1}$ at $z\sim15$. The MRD of Pop.~II BNSs peaks at $\mathcal{R}_{\rm BNS}(z) \sim 5\,\mathrm{Gpc}^{-3}\,\mathrm{yr}^{-1}$ around $z \sim 10$, dominating over the Pop.~III contribution up to that redshift. Despite the compactness of Pop.~III stars during their evolution (Fig.~\ref{fig:r_evol}), Pop.~II stars produce a MRD at least one order of magnitude higher than the Pop.~III one. This is mainly due to the impact of the star formation rate density models on the MRDs. Also in this case, configurations with initial orbital parameters from \cite{sb2013} produce smaller MRDs with respect to those with initial distributions from \cite{sana2012}.

We explored the impact of the initial conditions on the distributions of the primary mass, $m_1$ (Fig.~\ref{fig:bhns_m1_all}), and mass ratio, $q$ (Fig.~\ref{fig:bhns_q_all}), of BHNS mergers from Pop.~III stars. We find that all simulated models produce merging BHNSs with $6\lesssim m_1 \lesssim 50\,\msun$, except for one (log3), which yields BHNS mergers with primary masses of up to $400\,\msun$. Models with initial orbital parameters from \cite{sb2013} create preferentially merging systems with $m_1>15\,\msun$, while models with initial orbital parameters from \cite{sana2012} yield flatter $m_1$ distributions with a peak around $22\,\msun$. The impact of the initial orbital configuration is milder for Pop.~II stars, because of their largest expansion during their evolution.

We computed the detectability-conditioned Bayes factors \citep{mould2023} for three GW signals detected by LVK, classified either as BHNS mergers or as compact mergers with a secondary mass in the lower-mass gap ($2-5\,\msun$, \citealp{fryer2012}). In our analysis, we compared each simulated Pop.~III model to the fiducial Pop.~II model (see Table~\ref{table:odds}). GW191219 stands out as favoring a Pop.~III origin, particularly for models that predict mergers with extreme mass ratios. In contrast, GW signals with \(m_1 < 20\,\msun\) favor a Pop.~II origin.

In summary, our results indicate that, although Pop.~III stars are more compact than Pop.~II stars, their lower star formation rate density leads to MRDs of BHNSs and BNSs that are at least one order of magnitude lower than for BHNSs and BNSs born from Pop.~II binaries. This trend depends on the chosen models for the star formation rate density of Pop.~III \citep{hartwig2022} and Pop.~II stars \citep{hartwig2024, liu2025}. Finally, we find that Pop.~III BHNS mergers can involve massive BHs, and provide a possible formation channel for BHNS mergers with BH mass $>20\,\msun$ detected in the local Universe. The implications of our results for detection rates with second- and third-generation interferometers, as well as the effects of rotation and of different SN models and natal kick prescriptions on the component masses of BHNS mergers, will be explored in future works.

\begin{acknowledgements}
We thank the anonymous referee for their useful comments and suggestions, which helped us improve this manuscript.
MMa, GC, GI, and FS acknowledge financial support from the European Research Council for the ERC Consolidator grant DEMOBLACK, under contract no. 770017. MMa, RSK and BL also acknowledge financial support from the German Excellence Strategy via the Heidelberg Cluster of Excellence (EXC 2181 - 390900948) STRUCTURES. MB and MMa also acknowledge support from the PRIN grant METE under the contract no. 2020KB33TP. FS acknowledges financial support from the AHEAD2020 project (grant agreement n. 871158). 
MAS acknowledges funding from the European Union’s Horizon 2020 research and innovation programme under the Marie Skłodowska-Curie grant agreement No.~101025436 (project GRACE-BH) and from the MERAC Foundation. 

GC acknowledges partial financial support from European Union—Next Generation EU, Mission 4, Component 2, CUP: C93C24004920006, project ‘FIRES'.

GI is supported by a fellowship grant from the la Caixa Foundation (ID 100010434), code LCF/BQ/PI24/12040020. GI also acknowledges financial support under the National Recovery and Resilience Plan (NRRP), Mission 4, Component 2, Investment 1.4, - Call for tender No. 3138 of 18/12/2021 of Italian Ministry of University and Research funded by the European Union – NextGenerationEU.

MMo is supported by the LIGO Laboratory through the National Science Foundation awards PHY-1764464 and PHY-2309200.

RSK acknowledges financial support from the ERC via Synergy Grant "ECOGAL" (project ID 855130) and from the German Ministry for Economic Affairs and Climate Action in project "MAINN" (funding ID 50OO2206).  RSK also thanks the 2024/25 Class of Radcliffe Fellows for highly interesting and stimulating discussions. 

The authors acknowledge support by the state of Baden-W\"urttemberg through bwHPC and the German Research Foundation (DFG) through grants INST 35/1597-1 FUGG and INST 35/1503-1 FUGG.

We used \textsc{sevn} (\url{https://gitlab.com/sevncodes/sevn}) to generate our catalogs \citep{spera2019,mapelli2020,sevn2023}. We used \textsc{cosmo}$\mathcal{R}$\textsc{ate} (\url{https://gitlab.com/Filippo.santoliquido/cosmo_rate_public}) to generate our merger rate densities and convolve our catalogs of mergers with the star formation history \citep{santoliquido2020, santoliquido2021, santoliquido2023}.

We obtained our star formation rate densities from \textsc{a-sloth} (\url{https://gitlab.com/thartwig/asloth}). 

This research made use of \textsc{NumPy} \citep{harris2020}, \textsc{SciPy} \citep{scipy2020}, \textsc{Pandas} \citep{pandas2020}. For the plots we used \textsc{Matplotlib} \citep{hunter2007}.
\end{acknowledgements}

\bibliographystyle{aa} 
\bibliography{bibliography.bib} 

\begin{appendix}
\section{Initial conditions for Pop.~III binary stars}\label{ap:isolated}
In the following, we briefly present the initial distributions of primary mass, mass ratio, orbital period and eccentricity originally described by \cite{costa2023} and listed in Table~\ref{table:ic_costa}. 

\subsection{Initial mass functions}
We consider four different IMFs for the primary mass:
\begin{enumerate}[(i)]
  \setlength\itemsep{0.3em}
  \item A \cite{kroupa2001} IMF $\xi(m_{\mathrm{ZAMS}})\propto m_{\mathrm{ZAMS}}^{-2.3}$, which is commonly adopted for Pop.~II stars. With respect to the canonical IMF, here we consider a single slope since $m_{\textrm{min}}\geq0.5\,\msun$.
  \item A \cite{larson1998} IMF $\xi(m_{\mathrm{ZAMS}})\propto m_{\mathrm{ZAMS}}^{-2.35}\exp{\left(-m_{\mathrm{cut1}}/m_{\mathrm{ZAMS}}\right)}$, with $m_{\mathrm{cut1}}=20\,\msun$.
  \item A flat log distribution  $\xi(m_{\mathrm{ZAMS}})\propto m_{\mathrm{ZAMS}}^{-1}$ (\citealp{sb2013}; \citealp{hirano2015, susa2014, wollenberg2020, chon2021, tanikawa2021, jaura2022, prole2022}). 
  \item A top heavy distribution (\citealp{sb2013}; \citealp{jaacks2019, liu2020}), $\xi(m_{\mathrm{ZAMS}})\propto m_{\mathrm{ZAMS}}^{-0.17}\exp{\left(-m_{\mathrm{cut2}}^2/m_{\mathrm{ZAMS}}^2\right)}$, where $m_{\mathrm{cut2}}=20\,M_{\odot}$.
\end{enumerate}

\subsection{Mass ratio and secondary mass}
We derive the secondary mass of the ZAMS star with three different prescriptions:
\begin{enumerate}[(i)]
  \setlength\itemsep{0.3em}
  \item The $q$ distribution from \cite{sana2012}, $\xi(q)\propto q^{-0.1}$, with $q\in[0.1,1]$, which fits the mass ratios of O- and B-type binaries in the local Universe; here we set $m_{\mathrm{ZAMS,2}}\geq2.2\,\msun$.
  \item A "sorted" distribution: we draw the ZAMS mass of the entire stellar population from the chosen IMF. Then, we randomly pair two stars, enforcing $m_{\mathrm{ZAMS,2}}\leq m_{\mathrm{ZAMS,1}}$.
  \item The $q$ distribution from \citet{sb2013}, $\xi(q)\propto q^{-0.55}$, with $q\in[0.1,1]$. This distribution was obtained by fitting Pop.~III stars generated in cosmological simulations. Here we assume $m_{\mathrm{ZAMS,2}}\geq2.2\,\msun$.
\end{enumerate}

\subsection{Orbital period}
We assume two distributions for the initial orbital period $P$.
\begin{enumerate}[(i)]
  \setlength\itemsep{0.3em}
  \item $\xi(\Pi)\propto \Pi^{-0.55}$, with $\Pi=\log{(P/\mathrm{day})}\in[0.15,5.5]$ from \cite{sana2012}. 
  \item $\xi(\Pi)\propto \exp{\left[-\left(\Pi-\mu\right)^2/ \left(2\sigma^2\right)\right]}$, 
  that is a Gaussian distribution with $\mu=5.5$ and $\sigma=0.85$ \citep{sb2013}. 
\end{enumerate}

\subsection{Eccentricity}
We draw the orbital eccentricity $e$ from two distributions.
\begin{enumerate}[(i)]
  \setlength\itemsep{0.3em}
  \item $\xi(e)\propto e^{-0.42}$ with $e\in[0,1)$ \citep{sana2012}. 
  \item A thermal distribution $\xi(e)\propto e$ with $e\in[0,1)$ \citep{heggie1975,tanikawa2021, hartwig2016, kinugawa2014}, 
  which favors highly eccentric systems. As has been shown by \cite{park2021,park2023}, Pop.~III binaries form preferentially with high orbital eccentricity. 
\end{enumerate}

\section{BHNS and BNS MRD of Pop.~II-I stars}\label{app:sevn}
In Figs.~\ref{fig:mrd_bhns} and \ref{fig:mrd_bns}, we compare the MRDs of BHNSs and BNSs from Pop.~III and Pop.~II stars with those from Pop.~II-I. These MRDs were derived by \cite{sevn2023} using a fiducial configuration. Below, we summarize the main properties of this model and describe the computation of the MRD. For further details, we refer the reader to \cite{sevn2023}.

One million binary star systems were simulated at 15 different metallicities, $Z \in [10^{-4},\, 3\times10^{-2}]$. The initial conditions for these simulations match those of the kro1 model, except for the mass range considered. In \cite{sevn2023}, the primary mass $m_{\rm ZAMS,1}$ is drawn from a Kroupa IMF \citep{kroupa2001} in the range $[5,150]\,\msun$, as opposed to $[2,600]\,\msun$ in kro1. The initial distributions of mass ratio and orbital period follow \cite{sana2012}. The minimum secondary mass is set to $m_{\rm ZAMS,2}=2.2\,\msun$. The treatment of mass transfer, compact object formation, SNe, pair-instability SNe, and natal kicks is the same as described in Sec.~\ref{methods}. 

The resulting BHNS and BNS merger catalogs are used as input for \textsc{cosmo}$\mathcal{R}$\textsc{ate}. To compute the MRD of these systems, we use Eq.~\ref{eq:mrd} from Sec.~\ref{sec:mrd_methods}, assuming the star formation rate density $\psi(z)$ from \cite{madau2017}:
\begin{equation}
    \psi(z)=a\,\frac{(1+z)^b}{1+[(1+z)/c]^d}\,\rm [\msun\,yr^{-1}\,Mpc^{-3}],
\end{equation}
where $a=0.01\,\msun \rm\,yr^{-1}\,Mpc^{-3}$ (assuming a Kroupa IMF; \citealp{kroupa2001}), $b=2.6$, $c=3.2$, and $d=6.2$.

We also modify Eq.~\ref{eq:f_mrd} as follows:
\begin{equation}
\mathcal{F}(z',z,Z) = \frac{1}{M_{\ast}(z',Z)}\,\frac{{\rm{d}}\mathcal{N}(z',z,Z)}{{\rm{d}}t(z)}\,p(z',Z),
\end{equation}
where we assume the metallicity distribution $p(z',Z)$ from \cite{madau2017}:
\begin{equation}
    p(z',Z) = \frac{1}{\sqrt{2\pi\sigma_{\rm Z}^2}} \exp\left\{ -\frac{\left[\log(Z(z')/\mathrm{Z_{\odot}}) - \langle \log Z(z')/\mathrm{Z_{\odot}} \rangle\right]^2}{2\,\sigma_{\rm Z}^2} \right\}.
\end{equation}
Here, $\langle \log Z(z')/\mathrm{Z_{\odot}} \rangle = \log \langle Z(z')/\mathrm{Z_{\odot}} \rangle - \ln(10)\,\sigma_{\rm Z}^2/2$, and we adopt $\sigma_{\rm Z} = 0.2$ \citep{bouffanais2021}.

Finally, the total initial stellar mass is computed as $M_{*}(z',Z) = M_{\rm sim}/f_{\rm IMF}$, where $M_{\rm sim}$ is the total simulated mass and $f_{\rm IMF} = 0.285$, to account for the fact that the simulations only include stars with $m_{\rm ZAMS,1} > 5\,\msun$ and $m_{\rm ZAMS,2} > 2.2\,\msun$, while the Kroupa IMF is defined down to $0.1\,\msun$ \citep{kroupa2001}.

\section{Impact of Pop.~III star-formation rate models on BHNS and BNS MRDs}\label{app:sfrd_comparison}

In this Section, we evaluate how the choice of the Pop.~III star formation rate density  model affects the results presented in Sec.~\ref{sec:mrd_bhns} and Sec.~\ref{sec:mrd_bns}. Specifically, we focus on the Pop.~III models that yield the highest MRDs in Fig.~\ref{fig:mrd_bhns} and Fig.~\ref{fig:mrd_bns} - lar1 and kro1, respectively - and recompute the MRDs assuming star formation rate densities from \citet[][H22; fiducial model in the main body]{hartwig2022}, \citet[][LB20]{liu2020}, \citet[][SW20]{skinner2020}, \citet[][J19]{jaacks2019}, and \citet[][dS11]{desouza2011}. It should be noted that the optimistic star-formation rate density model presented by \cite{desouza2011} is in tension with the cosmological parameters from \cite{aghanim2020}. As a consequence, dS11 is a rescaled version of it by 0.3, following the formalism presented in \cite{kinugawa2019}.

\begin{figure}[ht]
    \centering
    \includegraphics[width=0.9\columnwidth]{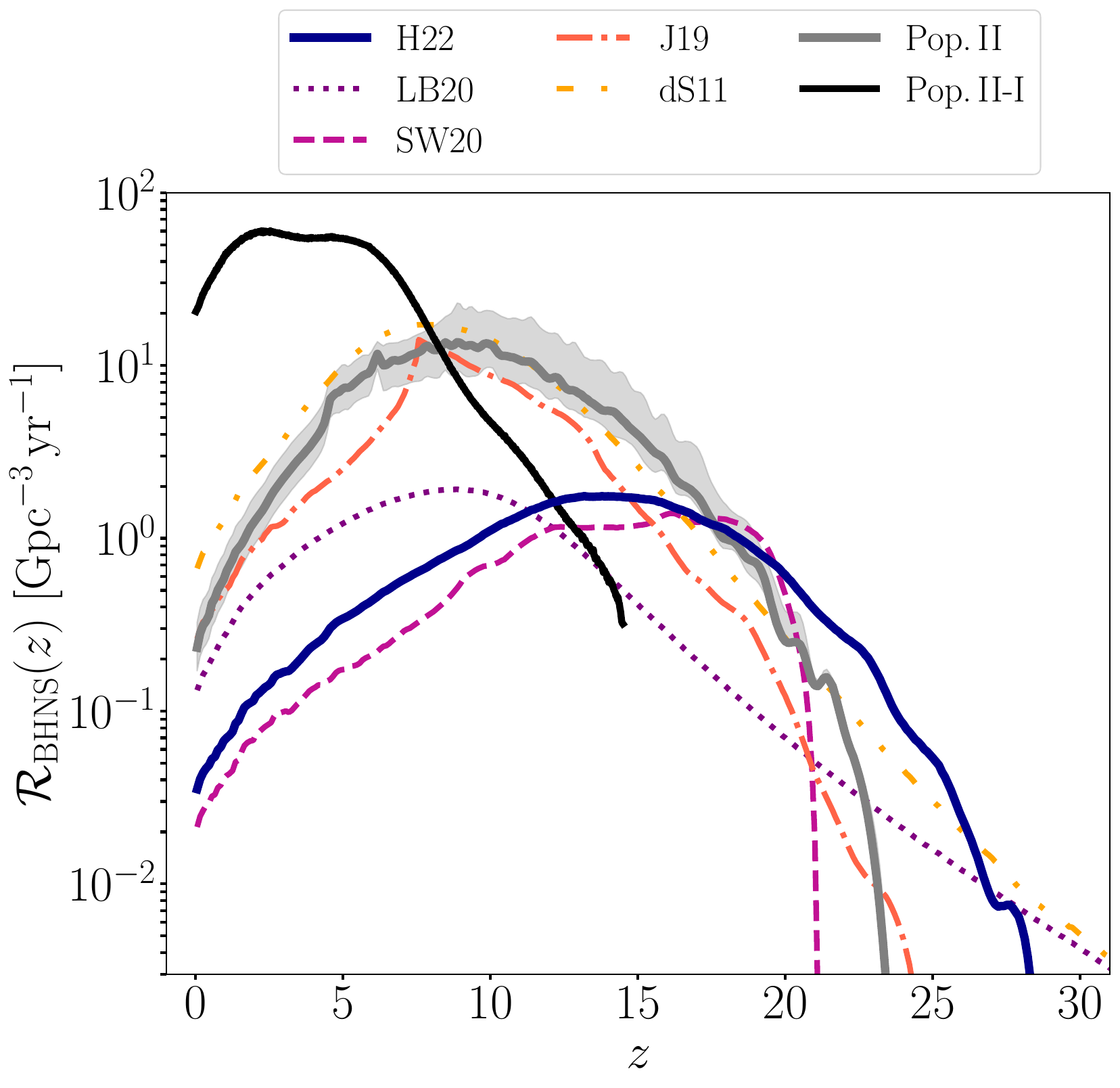}
    \caption{BHNS MRDs for the model producing the largest MRD in Fig.~\ref{fig:mrd_bhns} (lar1), considering different prescription for the Pop.~III star formation rate density. Thick blue line: \citet[][H22; fiducial model adopted in the main body]{hartwig2022}; violet dotted line: \citet[][LB20]{liu2020}; magenta dashed line: \citet[][SW20]{skinner2020}; orange dash-dotted line: \citet[][J19]{jaacks2019}; yellow dash-dot-dotted line: rescaled \citet[][dS11]{desouza2011}. As a comparison we also plot the BHNS MRD for early Pop.~II (gray line) and Pop.~II-I BHNSs (black line).}
     \label{fig:mrd_bhns_cfr}
\end{figure}

Figure~\ref{fig:mrd_bhns_cfr} shows the MRDs for BHNS systems. The H22, LB20, and J19 models yield MRDs that remain below the MRD from Pop.~II BHNSs up to high redshifts ($z \in \left[17,\,23\right]$). The SW20 model yields the lowest BHNS MRD of all the considered star-formation rate models for Pop.~III stars. In contrast, the dS11 model produces a BHNS MRD slightly higher than the Pop.~II one at $z \lesssim 7$  and then again from $z\sim22$ onward. Across the different models for the Pop.~III star formation rate evolution, the MRDs peak at $\mathcal{R}_{\rm BHNS} \in \left[ 2,\,20 \right]\,\rm{Gpc^{-3}\,yr^{-1}}$. At $z = 0$, the MRDs are consistently comparable to, or greater than, the fiducial value ($\sim 0.03\,\rm Gpc^{-3}\,yr^{-1}$), implying that the MRD of Pop.~III BHNSs with $m_1 > 20\,\msun$ is always equal to or higher than that of BHNSs from Pop.~II stars. This outcome highlights the robustness of our findings from the fiducial model.

\begin{figure}[ht]
    \centering
    \includegraphics[width=0.9\columnwidth]{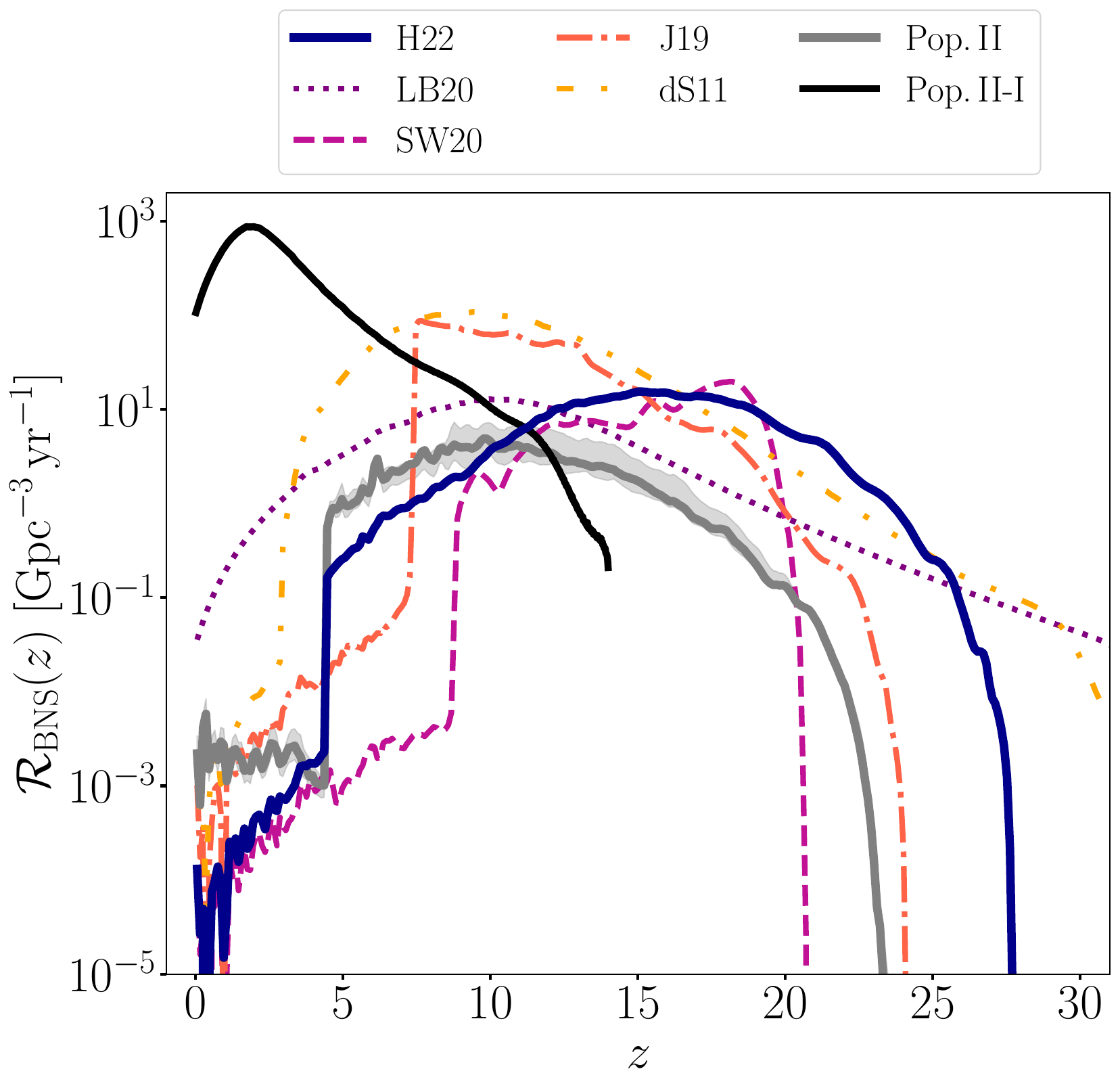}
    \caption{Same as Fig.~\ref{fig:mrd_bhns_cfr} but for the BNS MRDs, considering the Pop.~III model producing the largest MRD in Fig.~\ref{fig:mrd_bns} (kro1).}
     \label{fig:mrd_bns_cfr}
\end{figure}

Figure~\ref{fig:mrd_bns_cfr} shows the MRDs of BNSs for the various Pop.~III star formation rate density models. The H22, SW20, and J19 models exceed the Pop.~II MRD only from intermediate redshifts ($z \in \left[7,\,11\right]$) while LB20 and dS11 remain dominant throughout the entire redshift range. At $z = 0$, the LB20 model yields the highest MRD. As in Fig.~\ref{fig:mrd_bns}, a sharp decline is visible at $z < 5$, driven by the combination of short delay times and star formation rate densities that do not extend to lower redshifts; the sole exception is LB20, because \cite{liu2020} extrapolated their Pop.~III star formation rate density down to $z = 0$.

\section{Model selection}\label{ap:gw_comparison}
Here, we briefly summarize the method presented by \cite{mould2023}. According to Bayes' theorem, the posterior distribution of the measured properties of a GW event is given by
\begin{equation}
    p(\vec{\theta}|\vec{d},U) = \frac{\mathcal{L}(\vec{d}|\vec{\theta})\,\pi(\vec{\theta}|U)}{\mathcal{Z}(\vec{d}|U)},
\end{equation}
where \(\mathcal{L}(\vec{d}|\vec{\theta})\) is the likelihood that the vector of waveform parameters \(\vec{\theta}\) (e.g., masses, spins, redshifts, etc.) produces the observed data \(\vec{d}\), and \(\pi(\vec{\theta}|U)\) is the uninformative prior on the parameters. The normalization factor,
\begin{equation}
    \mathcal{Z}(\vec{d}|U)=\int \mathcal{L}(\vec{d}|\vec{\theta'})\,\pi(\vec{\theta'}|U)\,d\vec{\theta'}
\end{equation}
is the marginal likelihood that normalizes the posterior, representing the probability of observing the data given the chosen uninformative model \(U\). Both the prior \(\pi(\vec{\theta}|U)\) and the likelihood \(\mathcal{L}(\vec{d}|\vec{\theta})\) are conditioned on the uninformative model \(U\).

If we include an astrophysical model \(A\) as an informative prior, the posterior can be expressed as
\begin{equation}
    p(\vec{\theta}|\vec{d},A) = p(\vec{\theta}|\vec{d},U)\,\frac{\pi(\vec{\theta}|A)}{\pi(\vec{\theta}|U)}\,\frac{\mathcal{Z}(\vec{d}|U)}{\mathcal{Z}({\vec{d}|A)}}.
\end{equation}
Integrating over \(\vec{\theta}\) and rearranging, we obtain the Bayes factor, which quantifies the relative likelihood of the two prior models:
\begin{equation}
    \mathcal{B}_{A/U} = \frac{\mathcal{Z}(\vec{d}|A)}{\mathcal{Z}(\vec{d}|U)} = \int p(\vec{\theta}|\vec{d},U)\,\frac{\pi(\vec{\theta}|A)}{\pi(\vec{\theta}|U)}\,d\vec{\theta}.
\end{equation}
A value of \(\mathcal{B}_{A/U} > 1\) (\(< 1\)) implies that the astrophysical model \(A\) is more (less) likely than the uninformative prior \(U\) to have produced the observed data \(\vec{d}\). A decisive conclusion is typically reached when \(|\ln\mathcal{B}_{A/U}| \gtrsim 5\) \citep{jeffreys1939}. If we consider two different informative models, \(A\) and \(B\), the Bayes factor is simply \(\mathcal{B}_{A/B} = \mathcal{B}_{A/U} / \mathcal{B}_{B/U}\). 

When comparing two Bayesian models, we can also compute the posterior odds:
\begin{equation}
    \mathcal{O}_{A/B} = \frac{p(A|\vec{d})}{p(B|\vec{d})} = \frac{\pi(A)}{\pi(B)}\,\mathcal{B}_{A/B},
\label{eq:odds_b}
\end{equation}
which reduces to the Bayes factor \(\mathcal{B}_{A/B}\) when the priors \(\pi(A)\) and \(\pi(B)\) are assumed to be equal, i.e., when we cannot estimate the priors.

We can also account for selection effects associated with the detection of GW signals. When a GW signal is detected, it necessarily originates from the subset of the population that is observable. Therefore, we can generalize Eq.~\ref{eq:odds_b} by conditioning the evidence for an astrophysical model \( A \) on detectability (det). This leads to:
\begin{equation}
    \mathcal{Z}(\vec{d} | A, \mathrm{det}) = \frac{P(\mathrm{det} | \vec{d})\,\mathcal{Z}(\vec{d} | A)}{P(\mathrm{det} | A)},
\end{equation}
where the numerator captures information about the specific detected event, and the denominator accounts for the overall distribution of sources that the model predicts to be detectable. This detectability-conditioned evidence ensures that model predictions extending into undetectable regions of parameter space do not artificially improve agreement with observed data. Since $P(\mathrm{det} | \vec{d})=1$, we can write:
\begin{equation}
    \frac{\mathcal{Z}(\vec{d} | A, \mathrm{det})}{\mathcal{Z}(\vec{d}|U)}  = \frac{\mathcal{B}_{A/U}}{P(\mathrm{det} | A)}.
\end{equation}
Finally, we obtain the detection-weighted Bayes factor:
\begin{equation}
    \mathcal{D}_{A/B} = \frac{\mathcal{Z}(\vec{d} | A, \mathrm{det})}{\mathcal{Z}(\vec{d} | B, \mathrm{det})} = \frac{P(\mathrm{det}|B)}{P(\mathrm{det}|A)}\,\mathcal{B}_{A/B}.
\end{equation}

\end{appendix}

\end{document}